\documentclass[reprint, showpacs]{revtex4-1}
\usepackage{graphicx}
\usepackage{amsfonts}
\usepackage{natbib}

\newcommand{\be}{\begin{equation}}
\newcommand{\ee}{\end{equation}}
\newcommand{\bea}{\begin{eqnarray}}
\newcommand{\eea}{\end{eqnarray}}

\newcommand{\mA}{\left< A \right>}
\newcommand{\sA}{\left< A^2 \right>}

\begin{document}
\title{Classical and quantum mixed-type lemon billiards without stickiness}

\author{\v Crt Lozej}
\author{Dragan Lukman}
\author{Marko Robnik}

\affiliation{CAMTP - Center for Applied Mathematics and Theoretical
  Physics, University of Maribor, Mladinska 3, SI-2000 Maribor, Slovenia, European Union}


\date{\today}

\begin{abstract} The boundary of the lemon billiards is defined by
  the intersection of two circles of equal unit radius with the
  distance $2B$ between their centers, as introduced by
  Heller and Tomsovic in Phys. Today {\bf 46} 38 (1993). 
  This paper is a continuation of our recent paper
  on classical and quantum ergodic lemon billiard ($B=0.5$) with strong
  stickiness effects published in Phys. Rev. E {\bf 103} 012204 (2021).
  Here we study the classical and quantum lemon billiards,
  for the cases $B=0.42,\;0.55,\; 0.6$, which are mixed-type billiards without
  stickiness regions and thus serve as ideal examples of systems with
  simple divided phase space. The  classical phase portraits show
  the structure of one large chaotic sea with uniform
  chaoticity (no stickiness regions) surrounding a large regular island
  with almost no further substructure, being entirely covered by invariant tori.
  The boundary between the chaotic sea and the regular island is smooth,
  except for a few points. The classical transport time is
  estimated to be very short (just a few collisions), therefore
  the localization of the chaotic eigenstates is rather weak.
  The quantum states are characterized by the following {\em universal}
  properties of mixed-type systems without stickiness in the chaotic regions:
  (i) Using the Poincar\'e-Husimi (PH) functions the eigenstates are
  separated to the regular ones and chaotic ones.   The regular eigenenergies
  obey the Poissonian statistics, while the chaotic ones exhibit the
  Brody distribution with various values of the level repulsion
  exponent $\beta$, its value depending on the strength of the localization
  of the chaotic eigenstates. Consequently,  the total spectrum is
  well described by the Berry-Robnik-Brody (BRB) distribution.
  (ii) The entropy localization measure $A$ (also the normalized inverse
  participation ratio) has a bimodal universal distribution, where the
   narrow peak at small $A$  encompasses the regular eigenstates,
  theoretically well understood, while the peak
  at larger $A$ comprises the chaotic eigenstates, and is well described by the
  beta distribution. (iii) Thus the BRB energy level spacing
  distribution captures two effects: the {\em divided phase space}
  dictated by the classical Berry-Robnik parameter $\rho_2$
  measuring the relative  size of the largest chaotic region,
  in agreement with the Berry-Robnik picture,
  and the {\em localization of chaotic PH functions} characterized by the
  level repulsion (Brody) parameter $\beta$.
  (iv) Examination of the PH functions shows that they are supported
  either on the classical invariant tori in the regular islands
  or on the chaotic sea, where they are only weakly localized.  With
  increasing energy the localization of chaotic states
  decreases, as the PH functions tend towards uniform spreading over the
  classical chaotic region, and correspondingly $\beta$ tends to $1$.

\end{abstract}

\pacs{01.55.+b, 02.50.Cw, 02.60.Cb, 05.45.Pq, 05.45.Mt}

\maketitle

\section{Introduction}
\label{sec1}

This paper is a continuation of our previous recent
paper \cite{LLR2020} on classical
and quantum {\em ergodic} billiard ($B=0.5$) with strong stickiness effects,
from the family of lemon billiards introduced by Heller and Tomsovic
in 1993 \cite{HelTom1993}. In the present paper we study lemon billiards
\cite{Lozej2020} with the shape parameters $B=0.42,\; 0.55,\; 0.6$, which are
mixed-type billiards without
stickiness regions and thus serve as ideal examples of systems with
a simple divided phase space. The  classical phase portraits show
the structure of one large chaotic sea with uniform
chaoticity (no stickiness regions) surrounding a large regular island
with almost no further substructure, being entirely covered by invariant tori.
The boundary between the chaotic sea and the regular island is smooth,
except for a few points.

For a general introduction to the subjects in quantum chaos related to our
work the reader is referred to the previous paper \cite{LLR2020}.
Here we only mention the introductions to the general quantum chaos
in the books by St\"ockmann \cite{Stoe} and  Haake \cite{Haake}, and the
recent review papers on the stationary quantum chaos in generic
(mixed-type) systems \cite{Rob2016,Rob2020}.

The lemon billiards are explicitly defined in Sec. \ref{sec2}.
Recently the entire family of classical lemon billiards for
a dense set of about 4000 values of  $B\in [0.01,0.99975]$
(in steps of $dB=0.00025$) has been analyzed by Lozej \cite{Lozej2020}.
Based on this extensive work we were able to select the interesting
cases treated in this paper. The main purpose of the present paper
is the analysis of the selected quantum lemon billiards
$B=0.42,\; 0.55,\; 0.6$, with the following goals:
(i) to study the energy level statistics of the entire spectrum, as well as
separately of the regular and chaotic eigenstates, (ii) to calculate
the {\em Poincar\'e-Husimi} (PH) functions of the eigenstates,
analyze their structure in relationship
with the classical phase portrait, and to examine the quantum localization
of chaotic eigenstates and perform the analysis of their statistical
properties.

The main results are the following. The level spacing distribution
of the entire spectrum is well described by the Berry-Robnik (BR)
distribution, if the localization of chaotic eigenstates is absent,
and by Berry-Robnik-Brody (BRB) if the chaotic states are localized.
Moreover, the separated regular levels obey Poissonian statistics,
whilst the separated chaotic levels obey the Brody distribution
exhibiting at most weak localization reflected in $\beta \approx 1$. The
PH functions are found to be well supported either on invariant tori
in the regular island, or on the chaotic component. The entropy localization
measure of the PH functions, denoted by $A$, has a distribution
with two peaks: the peak at small $A$ is associated with the
regular PH functions, and is quantitatively well understood.
The second peak at larger value of $A$ is associated with the
chaotic PH functions and obeys the beta distribution, typical
for chaotic eigenstates associated with the classical chaotic
components with no stickiness (uniform "chaoticity"). With increasing
energy this beta distribution converges to a Dirac delta function
peaked at the maximum value of $A=A_0\approx 0.7$.

The paper is organized as follows. In Sec. \ref{sec2} we define
the lemon billiards and examine their classical dynamical properties. 
In Sec. \ref{sec3} we perform the statistical analysis of the energy
spectra. In Sec. \ref{sec4} we define and calculate
the Poincar\'e-Husimi (PH) functions and analyze their structure
in relationship with the classical phase portraits. In Sec. \ref{sec5}
we introduce the entropy localization measure of the chaotic
eigenstates and investigate its statistical properties.
In Sec. \ref{sec6} we discuss the results and present the
conclusions.

\section{The definition of the lemon billiards 
  and their classical dynamical properties}
\label{sec2}

\begin{figure}
 \begin{centering}
   \includegraphics[width=9cm]{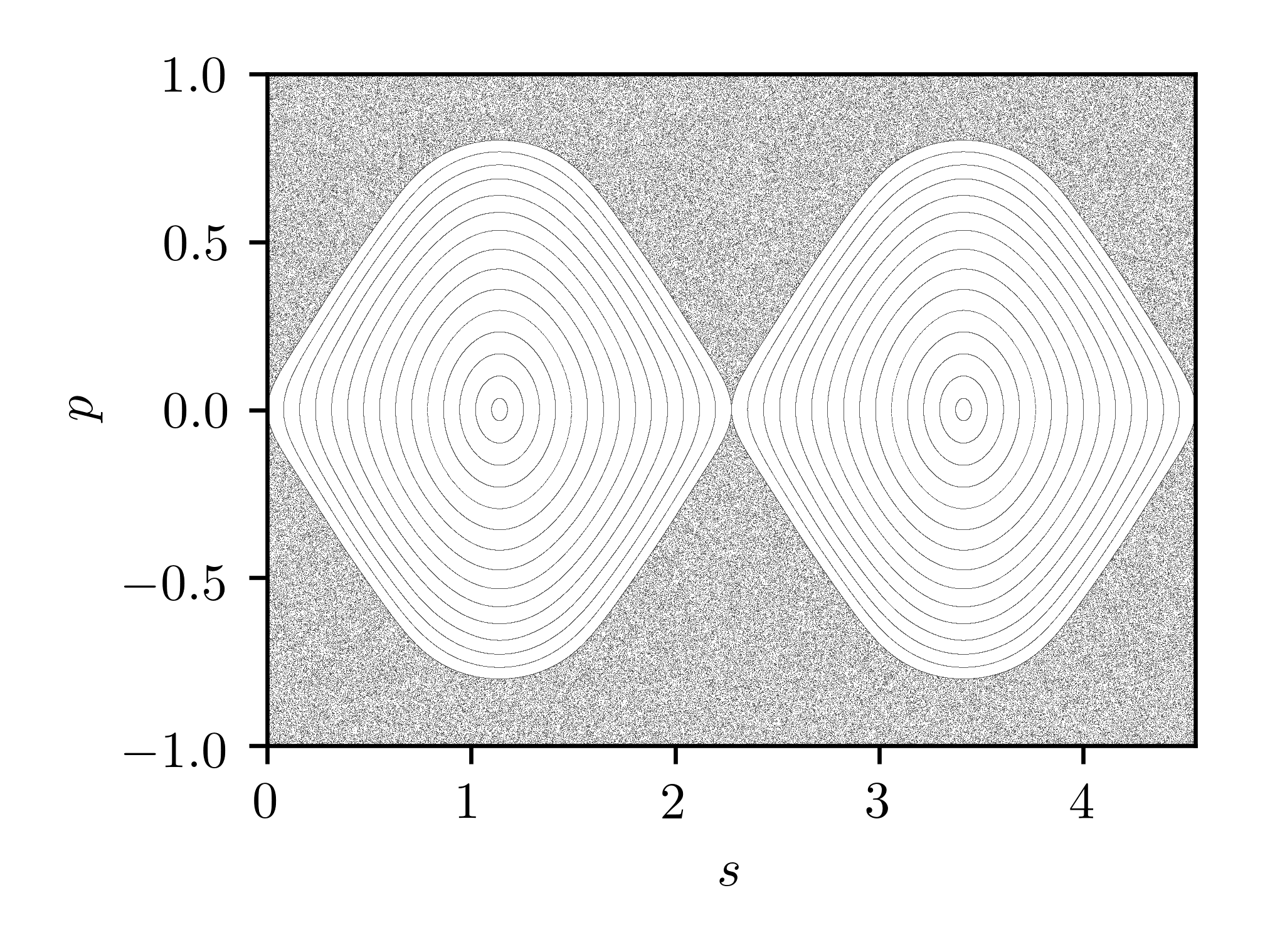}
   \par\end{centering}
 \caption{ The phase portrait of the lemon billiard $B=0.42$.
   The parameters are $\chi_c=0.4738$, and $\rho_2=0.4127$,
   $\rho_1=1-\rho_2=0.5873$. The label on the abscissa is $s$,
   while on the ordinate we have $p\in [-1,1]$.}
\label{fig1}
\end{figure}

The family of lemon billiards was introduced by Heller and Tomsovic
in 1993 \cite{HelTom1993}, and has been studied in a number of works
\cite{LMR1999,MHA2001,LMR2001,CMZZ2013,BZZ2016}, most recently
by Lozej \cite{Lozej2020} and Bunimovich et al \cite{BCPV2019},
and in our recent work \cite{LLR2020}.
The lemon billiard boundary is defined
by the intersection of two circles of equal unit radius with the
distance between their center $2B$ being less than their diameters
and $B\in (0,1)$, and is given by the following implicit equations
in Cartesian coordinates

\bea   \label{lemonB}
(x+B)^2 + y^2 =1, \;\;\; x > 0, \\  \nonumber
(x-B)^2 + y^2 =1, \;\;\; x < 0.
\eea
As usual we use the canonical variables to specify the location $s$
and the momentum component $p$ on the boundary at the collision point.
Namely the arclength $s$ counting in the mathematical positive sense
(counterclockwise) from the point $(x,y) = (0, -\sqrt{1-B^2})$ as the
origin, while $p$ is equal to the sine of the reflection angle $\theta$,
thus  $p = \sin \theta \in [-1,1]$, as $\theta \in [-\pi/2,\pi/2]$.
The bounce map $(s,p) \Rightarrow (s',p')$ is
area preserving as in all billiard systems \cite{Berry1981}.
Due to the two kinks the Lazutkin
invariant tori (related to the boundary glancing orbits) 
do not exist. The period-2 orbit connecting the centers of the two
circular arcs at the positions $(1-B,0)$ and $(-1+B,0)$ is always
stable (and therefore surrounded by a regular island) except for
the case $B=1/2$, where it is a marginaly unstable orbit (MUPO),
the case being ergodic and treated in our previous paper \cite{LLR2020}.
${\cal L}$ is the circumference of the entire billiard given by

\be \label{perimeter}
{\cal L} = 4 \arctan \sqrt{B^{-2} -1}.
\ee
The area ${\cal A} $ of the billiard is equal to

\be \label{area}
{\cal A} = 2 \arctan \sqrt{B^{-2} -1} - 2B\sqrt{1-B^2}.
\ee
The structure of the phase space is shown in Fig. \ref{fig1}
for the lemon billiard $B=0.42$.
The relative fraction of {\em the area} of the chaotic component
of the bounce map is $\chi_c=0.4738$, while the relative
fraction of {\em the phase space volume} of the same chaotic
component is $\rho_2=0.4127$ (which is the Berry-Robnik parameter).
The large regular island around the period-2 orbit is
densely covered by the invariant tori, 
with no visible thin chaotic layers, and the chaotic sea is
perfectly uniform, with no stickiness regions.

\begin{figure}
 \begin{centering}
    \includegraphics[width=9cm]{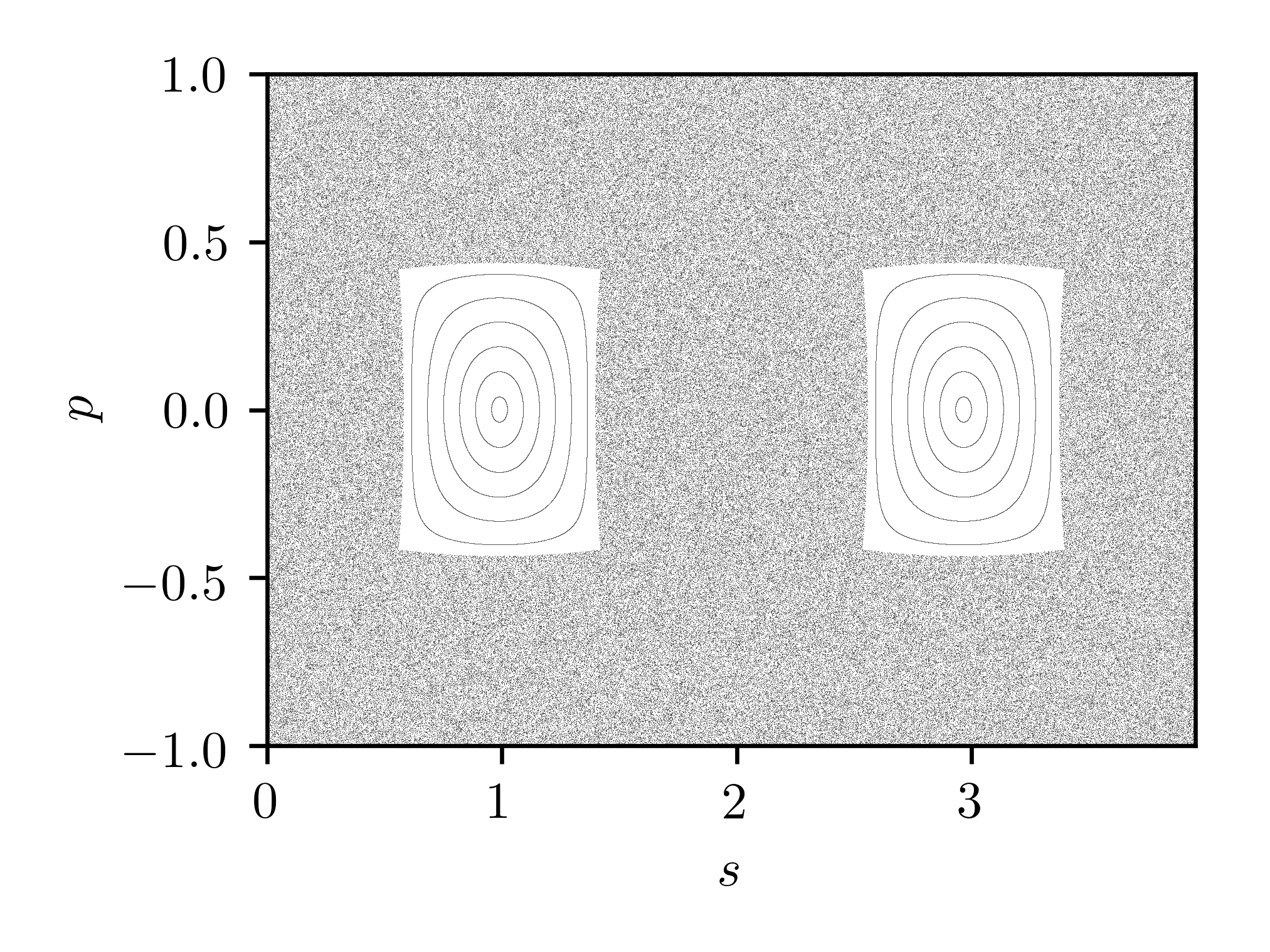}
   \par\end{centering}
 \caption{ As in Fig. \ref{fig1} but for  $B=0.55$.
   The parameters are $\chi_c=0.8204$, and $\rho_2=0.8076$,
   $\rho_1=1-\rho_2=0.1924$.}
\label{fig2}
\end{figure}

The structure of the phase space as shown in Fig. \ref{fig2}
for the lemon billiard $B=0.55$ is similar.
The relative fraction of the chaotic component
of the bounce map is $\chi_c=0.8204$, while the relative
fraction of the phase space volume of the same chaotic
component is $\rho_2=0.8076$.

\begin{figure}
 \begin{centering}
    \includegraphics[width=9cm]{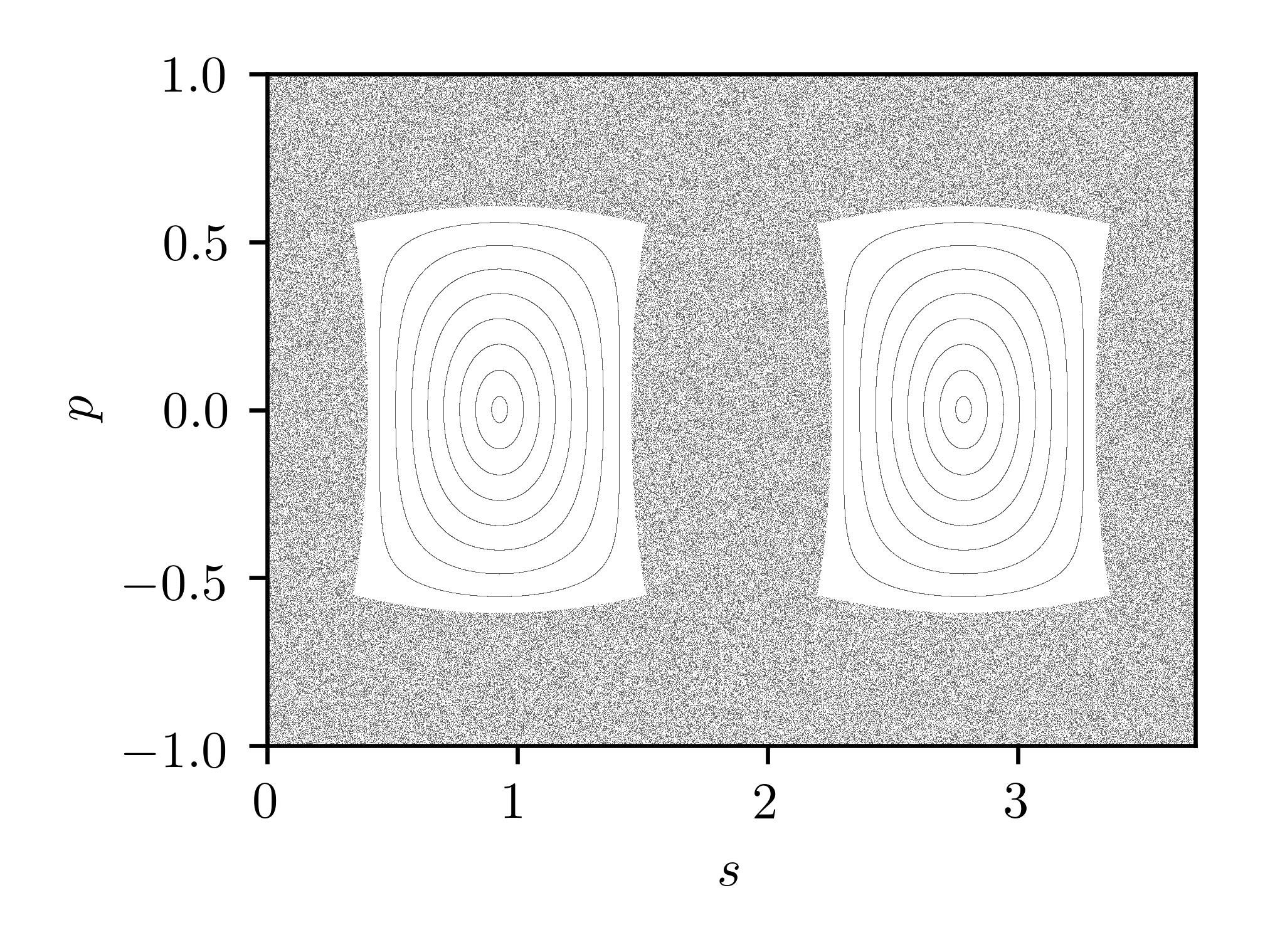}
   \par\end{centering}
 \caption{ As in Fig. \ref{fig1} but for  $B=0.6$.
   The parameters are $\chi_c=0.6545$, and $\rho_2=0.6338$,
   $\rho_1=1-\rho_2=0.3662$.}
\label{fig3}
\end{figure}

Also the lemon billiard $B=0.6$ is similar as shown in Fig. \ref{fig3}.
The relative fraction of the chaotic component
of the bounce map is $\chi_c=0.6545$, while the relative
fraction of the phase space volume of the same chaotic
component is $\rho_2=0.6338$.

We can conclude that the three cases $B=0.42,\;0.55,\; 0.6$ are
ideal to verify the Berry-Robnik picture of quantum billiards,
including the possible quantum localization, leading to the
Berry-Robnik-Brody level spacing distribution, and the universal
statistical properties of the localization measures, as there are no
stickiness effects, based on the results of the analysis of the
recurrence time statistics in Ref. \cite{Lozej2020},
unlike in the ergodic case $B=0.5$ studied
in Ref. \cite{LLR2020}.

\section{The energy level statistics}
\label{sec3}

We turn now to the quantum billiard ${\cal B}$ described
by the stationary Schr\"odinger equation, in the chosen units
($\hbar^2/2m=1$) given by the Helmholtz equation

\be \label{Helmholtz}
\Delta \psi + k^2 \psi =0
\ee
with the Dirichlet boundary conditions  $\psi|_{\partial {\cal B}}=0$.
The energy is $E=k^2$. 

The number of energy levels ${\cal N} (E)$ below $E=k^2$ is determined quite
accurately, especially at large energies, asymptotically exact,
by the celebrated Weyl formula (with perimeter corrections)
using the Dirichlet boundary conditions, namely

\be \label{WeylN}
{\cal N} (E) = \frac{{\cal A}\;E}{4\pi} - \frac{{\cal L}\;\sqrt{E}}{4 \pi} +c.c.,
\ee
where $c.c.$ are small constants determined by the corners and the
curvature of the billiard boundary.  Thus the density of levels
$d (E) = d{\cal N}/dE$ is equal to

\be \label{Weylrho}
d(E) = \frac{{\cal A}}{4\pi} - \frac{{\cal L}}{8 \pi \sqrt{E}}.
\ee
Our numerical method to solve the Helmholtz equation
is based on the Heller's plane wave
decomposition method and the Vergini-Saraceno scaling method
\cite{VerSar1995,LozejThesis}.
The numerical accuracy has been checked by the Weyl formula, to make sure
that we are neither losing levels nor getting too many due to the
double counting (distinguishing almost degenerate pairs from the
numerical pairs) in the overlapping energy intervals, and also by
the convergence test. The number of missing or too many levels was
never larger than 1 per 1000 levels (usually less than 10 per 10000 levels). 

Our billiard has two reflection symmetries, thus four symmetry classes:
even-even, even-odd, odd-even and odd-odd.  For the purpose of analyzing
the spectral statistics, the wavefunctions and the corresponding
PH functions, we have considered only the quarter billiard.

We have calculated the energy spectra for each billiard $B$ in nine
spectral stretches starting at $k_0=640$ in steps of 280, namely
$k_0=$ 640, 920, 1200, 1480, 1760,  2040, 2320, 2600, 2880, of various lengths,
from 1000 levels each up to 10000 levels each. We have used
the Weyl formula (\ref{WeylN}) for the spectral unfolding.

One of the most important statistical measures of the (unfolded)
energy spectra 
is the level spacing distribution $P(S)$. For integrable systems
we have Poissonian statistics and $P_P(S)= \exp(-S)$, while for
classical ergodic (fully chaotic) systems we have Wigner distribution
(Wigner surmise, which is 2-dim GOE formula), as an excellent
approximation for the GOE level spacing distribution ($\infty$-dim), 

\be \label{WPS}
P_W(S) = \frac{\pi S}{2} \exp \left(-\frac{\pi S^2}{4} \right).
\ee
There is a general very useful relationship, namely using the
gap probability  $E(S)$, which is the probability of having no
level on an arbitrary interval of length $S$.
The level spacing distribution $P(S)$ is in general equal to
the second derivative of the gap probability $P(S)=d^2{\cal E}(S)/dS^2$.

For the Poisson
statistics we have $E_P(S)=\exp(-S)$, while for the Wigner
distribution we find

\be \label{WGP}
{\cal E}_W(S) =  1 - {\rm erf} \left( \frac{\sqrt{\pi}S}{2} \right) 
 = {\rm erfc} \left( \frac{\sqrt{\pi}S}{2} \right).
\ee
In mixed-type systems we have typically one dominant chaotic
component with the relative density of levels $\rho_2$ (equal
to the relative fraction of the chaotic phase space volume),
while its complement is typically a regular component of relative
density $\rho_1=1-\rho_2$. If the regular and chaotic levels superimpose
statistically independent of each other, then obviously the
gap probability factorizes

\be \label{gapprob}
{\cal E}(S) = {\cal E}_{P}(\rho_1S) \; {\cal E}_{W}(\rho_2S),
\ee
and therefore in this case the level spacing distribution is
given by the Berry-Robnik formula \cite{BerRob1984}

\bea \label{BerryRobnik}
P_{BR}(S)  &=&   e^{-\rho_1 S}  e^{- \frac{\pi \rho_2^2 
S^2}{4}} \left( 2 \rho_1 \rho_2 + \frac{\pi \rho_2^3 S}{2} \right) \\ \nonumber
  &+&   e^{-\rho_1 S} \rho_1^2 {\rm erfc}
\left( \frac{\sqrt{\pi} \rho_2 S}{2} \right). 
\eea
The above statements are true provided the Heisenberg time is larger than
any classical transport time in the system \cite{Rob2020}. If this is not the
case, the chaotic eigenstates can be quantum (or dynamically) localized, which
implies localized chaotic PH functions to be introduced in the next
Sec. \ref{sec4}, and the level spacing distribution for such localized
chaotic eigenstates becomes (approximately) the well known Brody distribution,
\cite{Bro1973,Bro1981}, described by the following formula

\be \label{BrodyP}
P_B(S) = c S^{\beta} \exp \left( - d S^{\beta +1} \right), \;\;\; 
\ee
where by normalization of the total probability and the first moment we have

\be \label{Brodyab}
c = (\beta +1 ) d, \;\;\; d  = \left( \Gamma \left( \frac{\beta +2}{\beta +1}
 \right) \right)^{\beta +1}
\ee
with  $\Gamma (x)$ being the Gamma function. It interpolates the
exponential and Wigner distribution as $\beta$ goes from $0$ to $1$.
The important feature of the Brody distribution is the fractional
level repulsion effect, meaning the power law at small $S$,
$P(S)\propto S^{\beta}$. The corresponding gap probability is

\be \label{BrodyE}
{\cal E}_B(S)  =  \frac{1}{\gamma (\beta +1)  } 
  Q \left( \frac{1}{\beta +1}, \left( \gamma S \right)^{\beta +1} \right),
\ee
\\
where $\gamma=\Gamma \left(\frac{\beta +2}{\beta +1}\right)$ 
and $Q(a, x)$ is the incomplete Gamma function

\be \label{IGamma}
Q(a, x) = \int_x^{\infty} t^{a-1} e^{-t} dt.
\ee
Here the only parameter is $\beta$, the level repulsion exponent in
(\ref{BrodyP}), which measures the degree of localization of
the chaotic eigenstates: if the localization is maximally strong,
the eigenstates practically do not overlap in the phase space
(of the Wigner functions) and we find $\beta=0$ and Poissonian
distribution, while in the case of maximal extendedness (no localization)
we have $\beta=1$, and the GOE statistics of levels applies.
Thus, by replacing ${\cal E}_{W}(S)$ with ${\cal E}_B(S)$ we get the so-called
Berry-Robnik-Brody (BRB) distribution,
which generalizes the Berry-Robnik (BR) distribution such that the localization
effects are included \cite{BatRob2010}.
In this way the problem of describing the energy level statistics
is empirically solved. However, the theoretical derivation of
the Brody distribution for the localized chaotic states remains
an important open problem. Also, while the local behaviour at small
$S$, described by the power law  $P(S)\propto S^{\beta}$, is certainly
correct, the global feature of the Brody distribution is surely
approximate, although empirically very well founded.

In our present study the classical transport time of the billiards is
very short, therefore we expect $\beta \approx 1$, and the level spacing
distribution is almost Berry-Robnik (\ref{BerryRobnik}). Thus, in the level
statistics we do not detect large localization effects, but
accordingly, as we shall see, the PH functions still exhibit
some localization manifested by the entropy localization measure $A$.

\begin{figure}
 \begin{centering}
    \includegraphics[width=9cm]{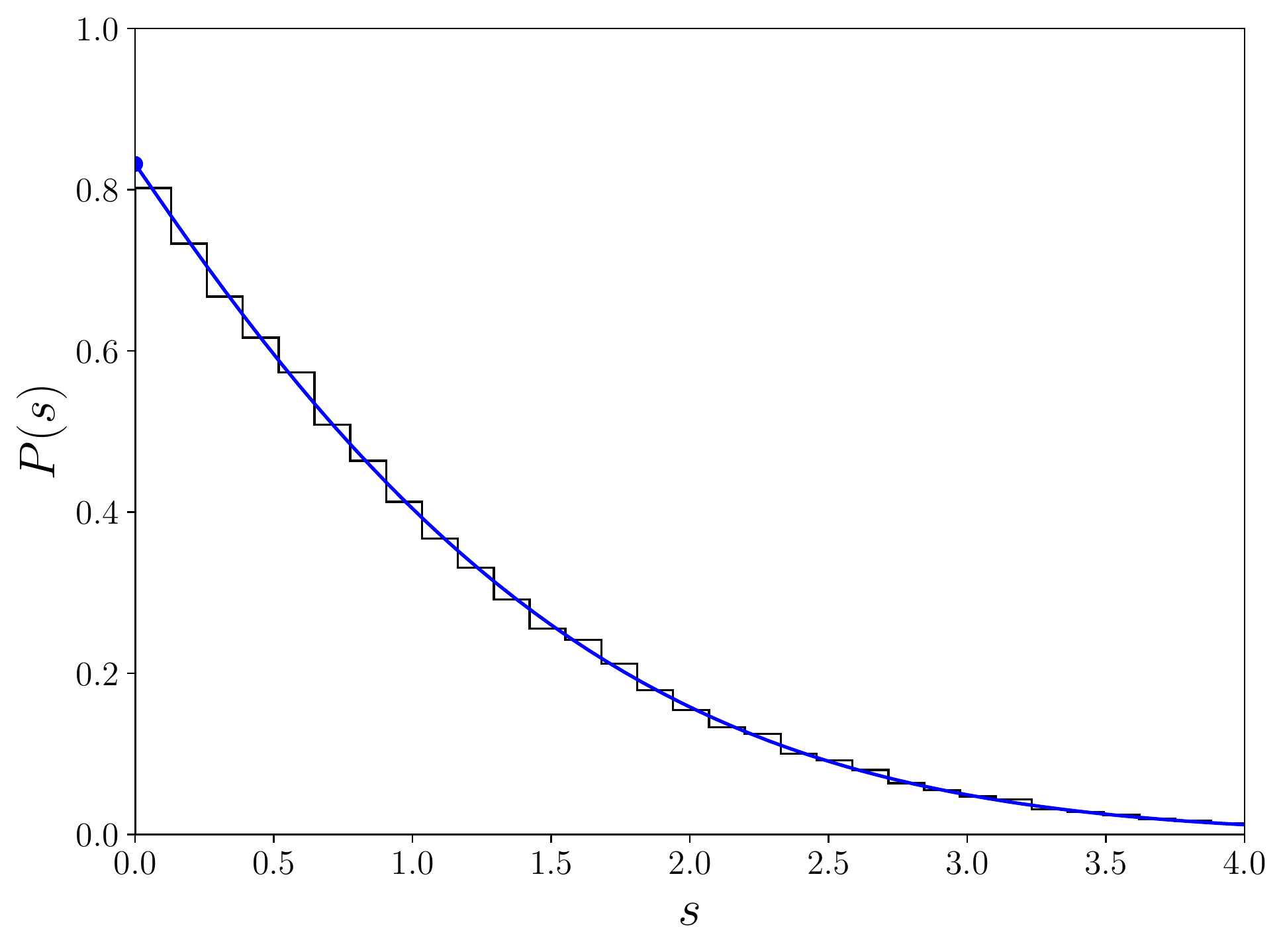}
   \par\end{centering}
 \caption{Level spacing distribution of the billiard $B=0.42$,
   for a spectral stretch starting at $k_0=2880$, comprising
   39841 levels of all parities.
   The classical parameter is $\rho_2=0.4127$,
   $\rho_1=1-\rho_2=0.5873$. The blue curve is the best fitting
   Berry-Robnik-Brody distribution. The thick dot designates the
 value of $P(S=0)$. The quantum $\rho_1=0.590$, and $\beta=0.941$.}
\label{fig4}
\end{figure}

\begin{figure}
 \begin{centering}
    \includegraphics[width=9cm]{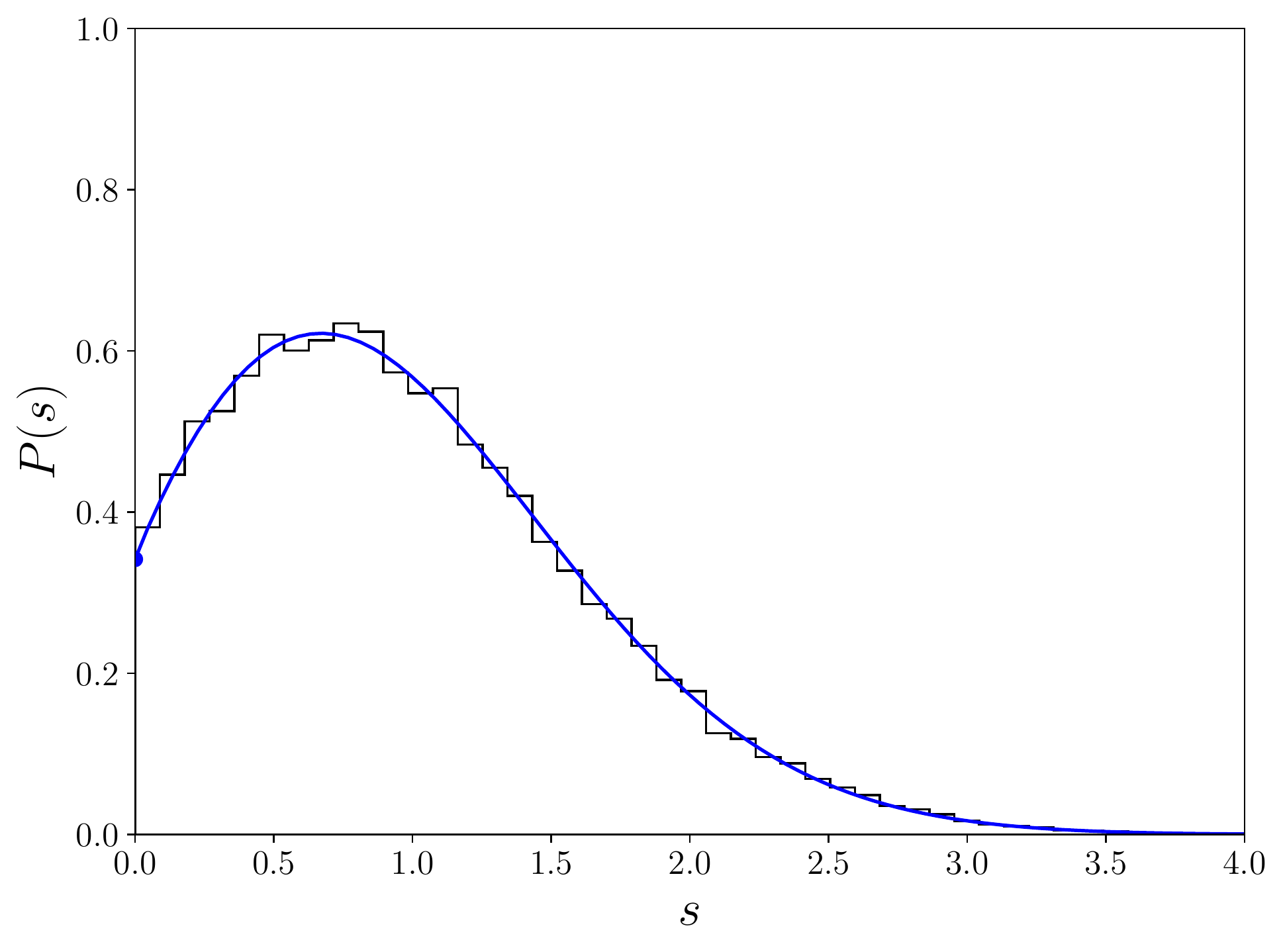}
   \par\end{centering}
 \caption{Level spacing distribution of the billiard $B=0.55$,
   for a spectral stretch starting at $k_0=2880$, comprising
   39874 levels of all parities.
   The classical parameter is $\rho_2=0.8076$,
   $\rho_1=1-\rho_2=0.1924$. The blue curve is the best fitting
   Berry-Robnik-Brody distribution. The thick dot designates the
 value of $P(S=0)$. The quantum $\rho_1=0.189$,  and $\beta=0.956$.}
\label{fig5}
\end{figure}

\begin{figure}
 \begin{centering}
    \includegraphics[width=9cm]{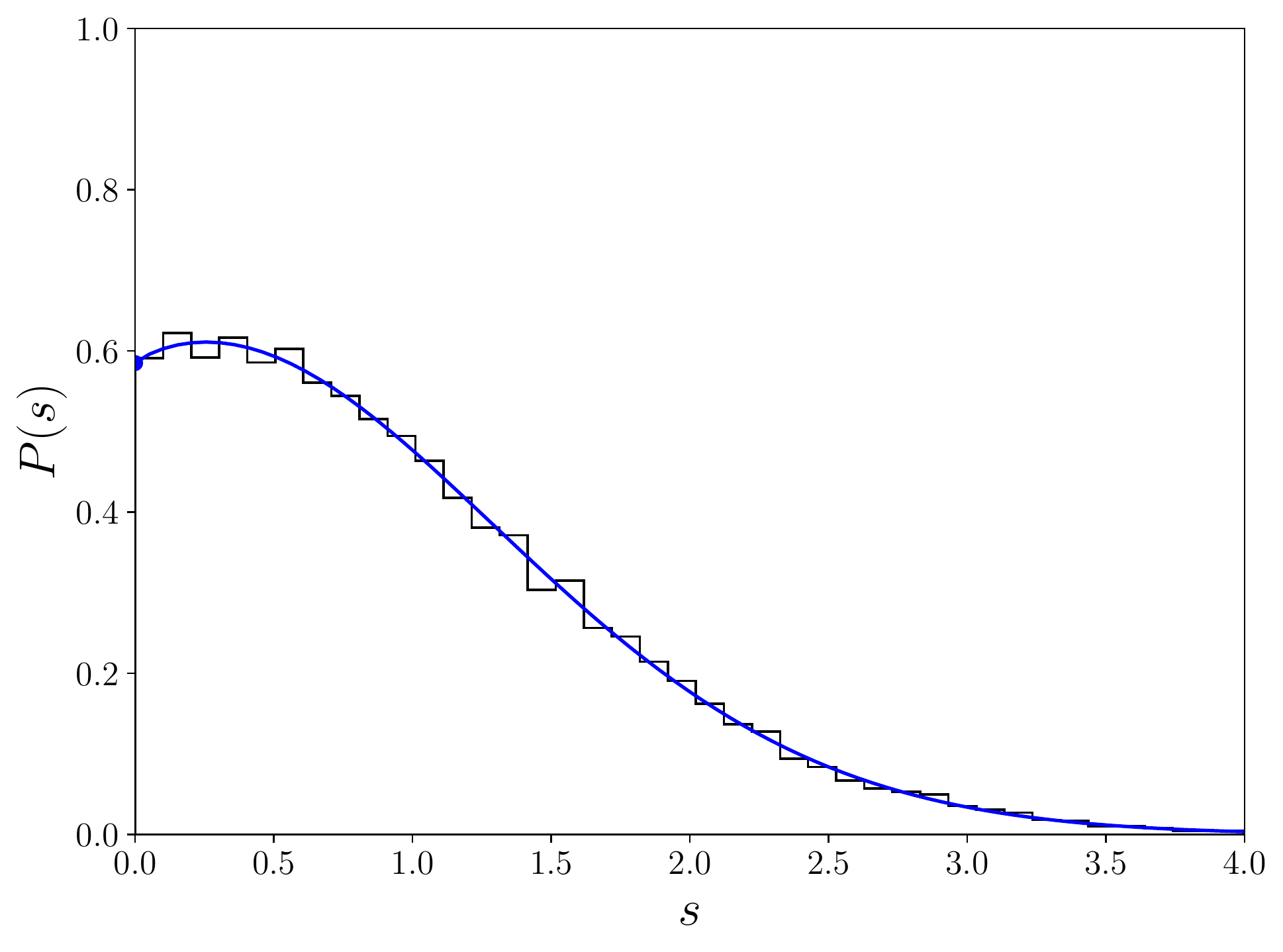}
   \par\end{centering}
 \caption{Level spacing distribution of the billiard $B=0.6$,
   for a spectral stretch starting at $k_0=2880$, comprising
   39899 levels of all parities.
   The classical parameter is $\rho_2=0.6338$,
   $\rho_1=1-\rho_2=0.3662$. The blue curve is the best fitting
   Berry-Robnik-Brody distribution. The thick dot designates the
 value of $P(S=0)$. The quantum $\rho_1=0.356$,  and $\beta=0.930$.}
\label{fig6}
\end{figure}

In Figs. \ref{fig4}-\ref{fig6}
we show the level spacing distributions $P(S)$ for
the three billiards $B=0.42,\;0.55,\;0.6$, respectively, for
the spectral stretches starting at $k_0=2880$, for all four parities
together (almost 40.000 levels),
with the best fitting BRB distribution. As we see the $\beta$ parameter
is very close to $1$, while the parameter $\rho_1$ is close to
its classical value. Thus the system exhibits the Berry-Robnik
picture with weak localization effects in the chaotic part
of the energy spectrum, well fitted by BRB distribution.
The picture is very similar for other values of $k_0$ listed above.

\begin{figure}
 \begin{centering}
    \includegraphics[width=9cm]{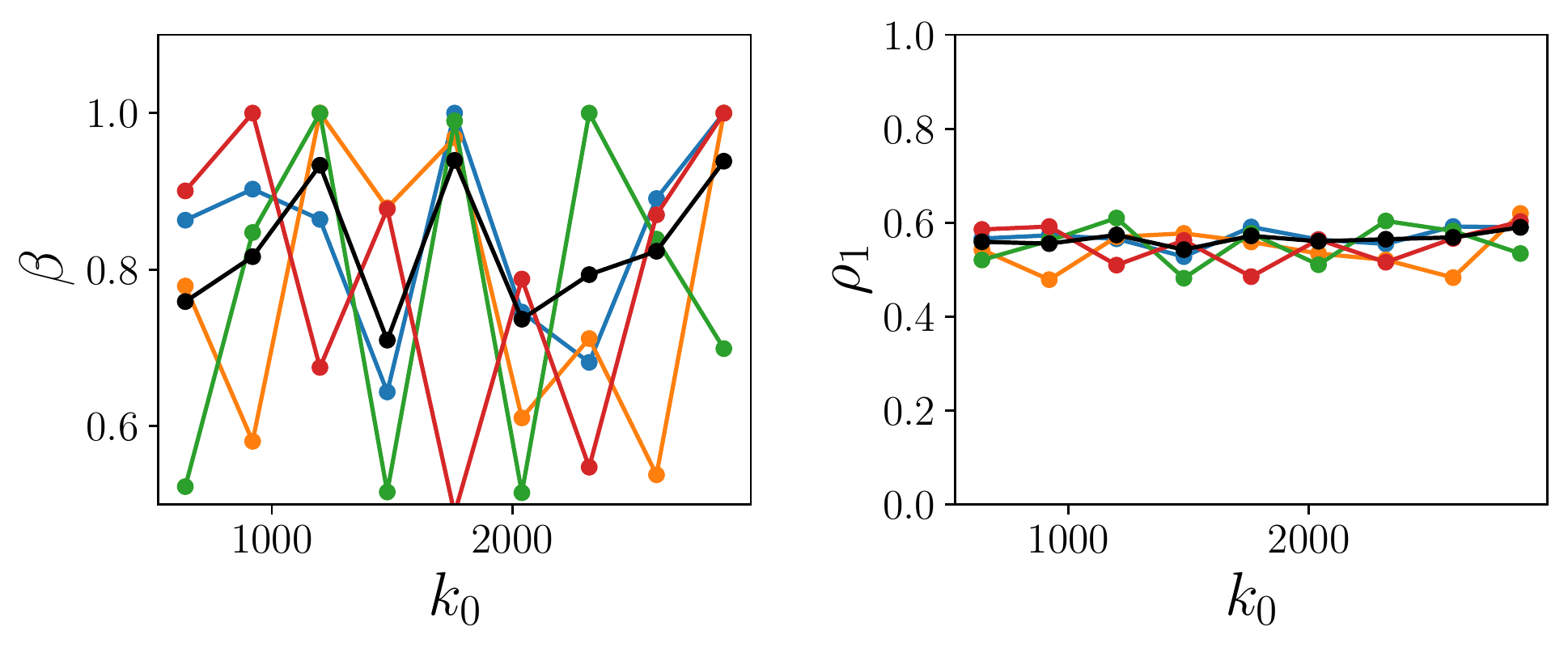}
   \par\end{centering}
 \caption{The billiard $B=0.42$, the best fitting $\beta$
   and best fitting $\rho_1$ vs. $k_0$, based on
   a spectral stretch  comprising
   about 10.000 levels for each parity, and about 40.000
   of all parities. The classical $\rho_1=1-\rho_2=0.5873$.
 For the legend see Figs. \ref{fig8} or \ref{fig9}.}
\label{fig7}
\end{figure}

\begin{figure}
 \begin{centering}
    \includegraphics[width=9cm]{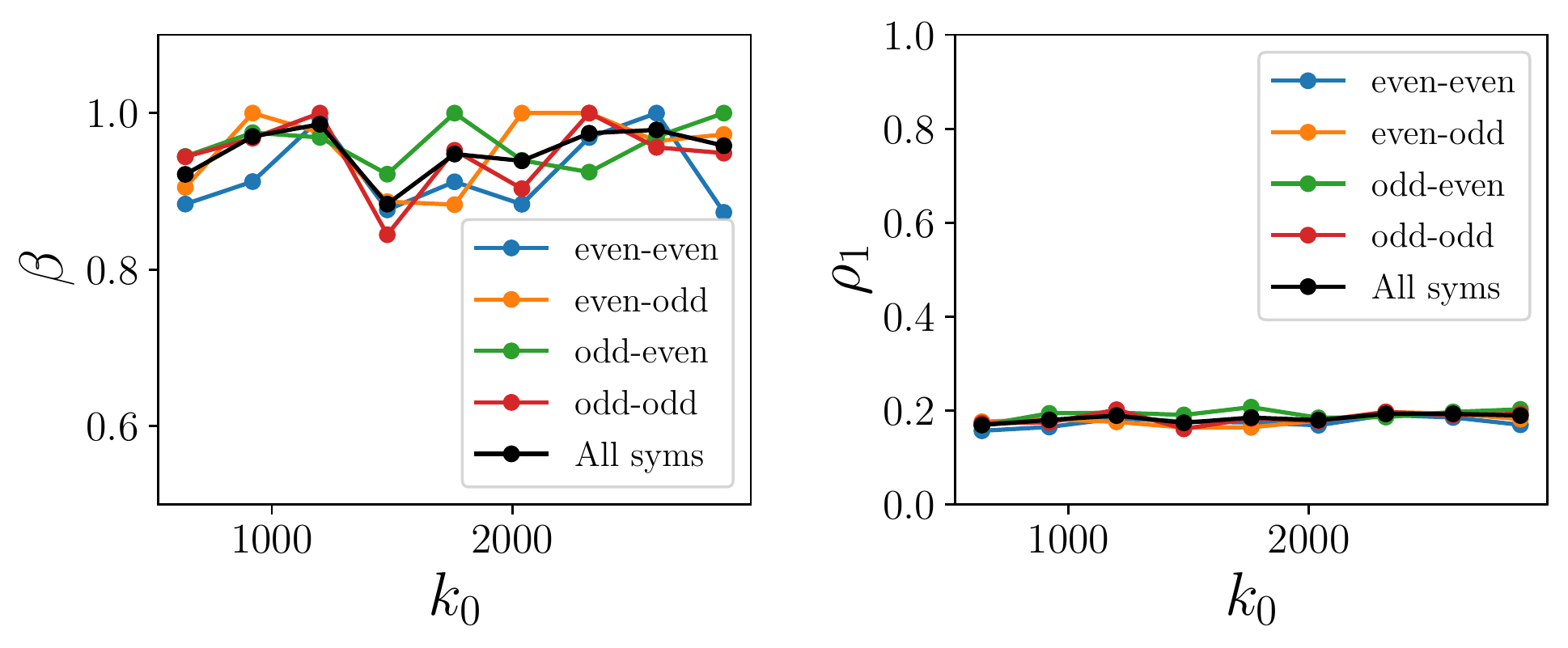}
   \par\end{centering}
 \caption{The billiard $B=0.55$, the best fitting $\beta$
   and best fitting $\rho_1$ vs. $k_0$, based on
   a spectral stretch  comprising
   about 10.000 levels for each parity, and about 40.000
   of all parities. The classical $\rho_1=1-\rho_2=0.1924$.}
\label{fig8}
\end{figure}

\begin{figure}
 \begin{centering}
    \includegraphics[width=9cm]{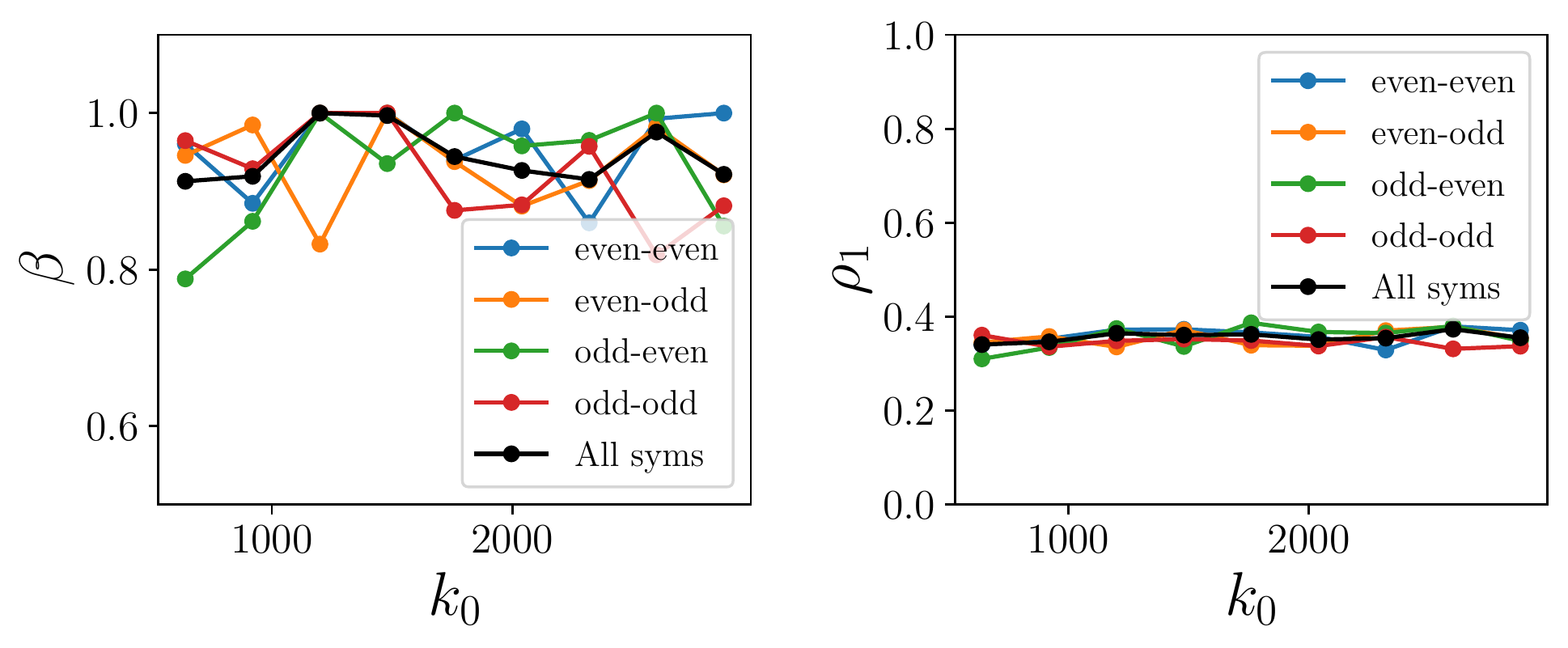}
   \par\end{centering}
 \caption{The billiard $B=0.6$, the best fitting $\beta$
   and best fitting $\rho_1$ vs. $k_0$, based on
   a spectral stretch  comprising
   about 10.000 levels for each parity, and about 40.000
   of all parities. The classical $\rho_1=1-\rho_2=0.3662$.}
\label{fig9}
\end{figure}

In Figs. \ref{fig7}-\ref{fig9}  we show the dependence of the
best fitting $\beta$ and $\rho_1$ values as functions of $k_0$,
from $640$ to $2880$, for all four parities separately and for
all of them taken together. We clearly see that the
quantum value of $\rho_1$ agrees very well with the classical value.
The value of $\beta$ is close to $1$, which indicates only
weak localization of the chaotic eigenstates.
The fluctuations of $\beta$ with $k_0$ are
the strongest in case $B=0.42$, and the smallest in case $B=0.55$.

In the next sections we shall treat the PH functions and based
on them will separate the regular and chaotic eigenstates,
using the entropy localization measure $A$, and will show
that the relative number of regular levels agrees again with
the classical value $\rho_1$, confirming the Berry-Robnik picture.


\section{The structure of Poincar\'e-Husimi functions}
\label{sec4}

Instead of studying the eigenstates by means of the wavefunctions
$\psi_m({\bf r})$ as solutions of the Helmholtz
equation (\ref{Helmholtz}) we define
the Poincar\'e-Husimi functions in the quantum phase space. PH
functions are a special case of Husimi functions \cite{Hus1940},
which are in turn Gaussian smoothed Wigner functions \cite{Wig1932}.
They are very natural for billiards. Following Tuale and Voros
\cite{TV1995} and B\"acker et al \cite{Baecker2004} we define the properly
${\cal L}$-periodized coherent states
centered at $(q,p)$, as follows

\bea \label{coherent}
c_{(q,p),k} (s) & =  & \sum_{m\in {\bf Z}} 
\exp \{ i\,k\,p\,(s-q+m\;{\cal L})\}  \times \\ \nonumber
 & \exp & \left(-\frac{k}{2}(s-q+m\;{\cal L})^2\right). 
\eea
The Poincar\'e-Husimi function is then defined as the absolute square
of the projection of the boundary function $u_m(s)$ onto the coherent
state, namely

\be \label{Husfun}
H_m(q,p) = \left| \oint c_{(q,p),k_m} (s)\;
u_m(s)\; ds \right|^2.
\ee
where $u_m(s)$ is the boundary function, that is the normal derivative
of the eigenfunction of the $m$-th state
$\psi(_m{\bf r})$ on the boundary at point $s$,

\be  \label{BF}
u_m(s) = {\bf n}\cdot \nabla_{{\bf r}} \psi_m \left({\bf r}(s)\right).
\ee
Here ${\bf n}$ is a unit outward normal vector to the
boundary at point ${\bf r}(s)$.
The boundary function satisfies an integral equation and
also uniquely determines the value of the wavefunction $\psi_m ({\bf r})$
at any interior point ${\bf r}$ inside the billiard ${\cal B}$.

According to the principle of uniform semiclassical condensation (PUSC) of
the Wigner functions and Husimi functions \cite{Rob1998,Rob2020}
the PH functions are expected to condensate (collapse) in the semiclassical
limit either on an invariant torus or on the chaotic component in the
classical phase space.

\begin{figure}
 \begin{centering}
    \includegraphics[width=9cm]{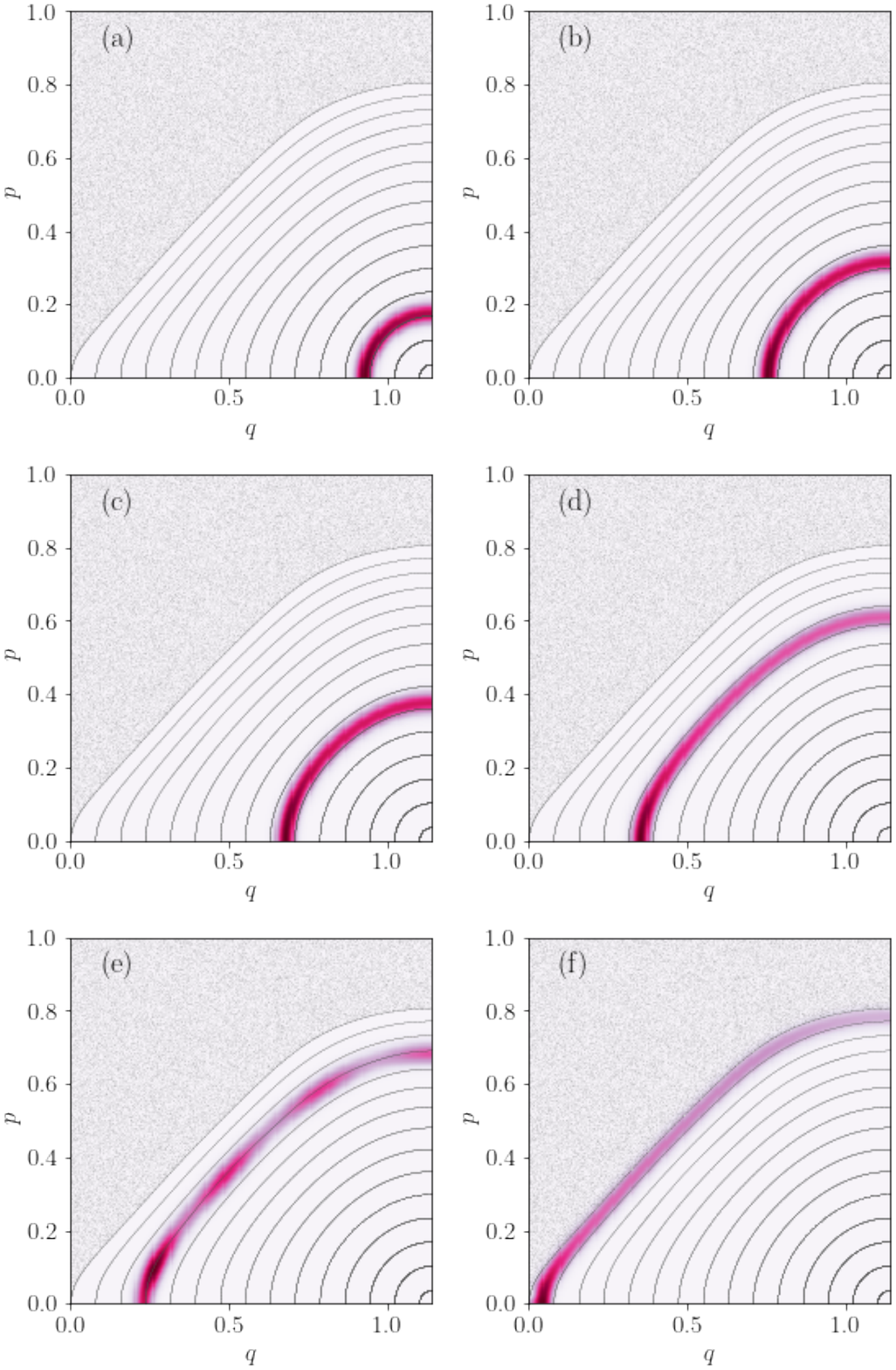}
   \par\end{centering}
 \caption{The billiard $B=0.42$, typical regular PH functions.
   We show only 1/8 of the phase space, due to the symmetries,
   for the even-even parity. The data $(k,A)$ are as follows from
   (a) to (f): (1317.630684, 0.021), (1303.594952, 0.038), (1289.244533, 0.046),
 (1279.769267, 0.074), (1220.262865, 0.084), (1308.380980, 0.095).}
\label{fig10}
\end{figure}

\begin{figure}
 \begin{centering}
    \includegraphics[width=9cm]{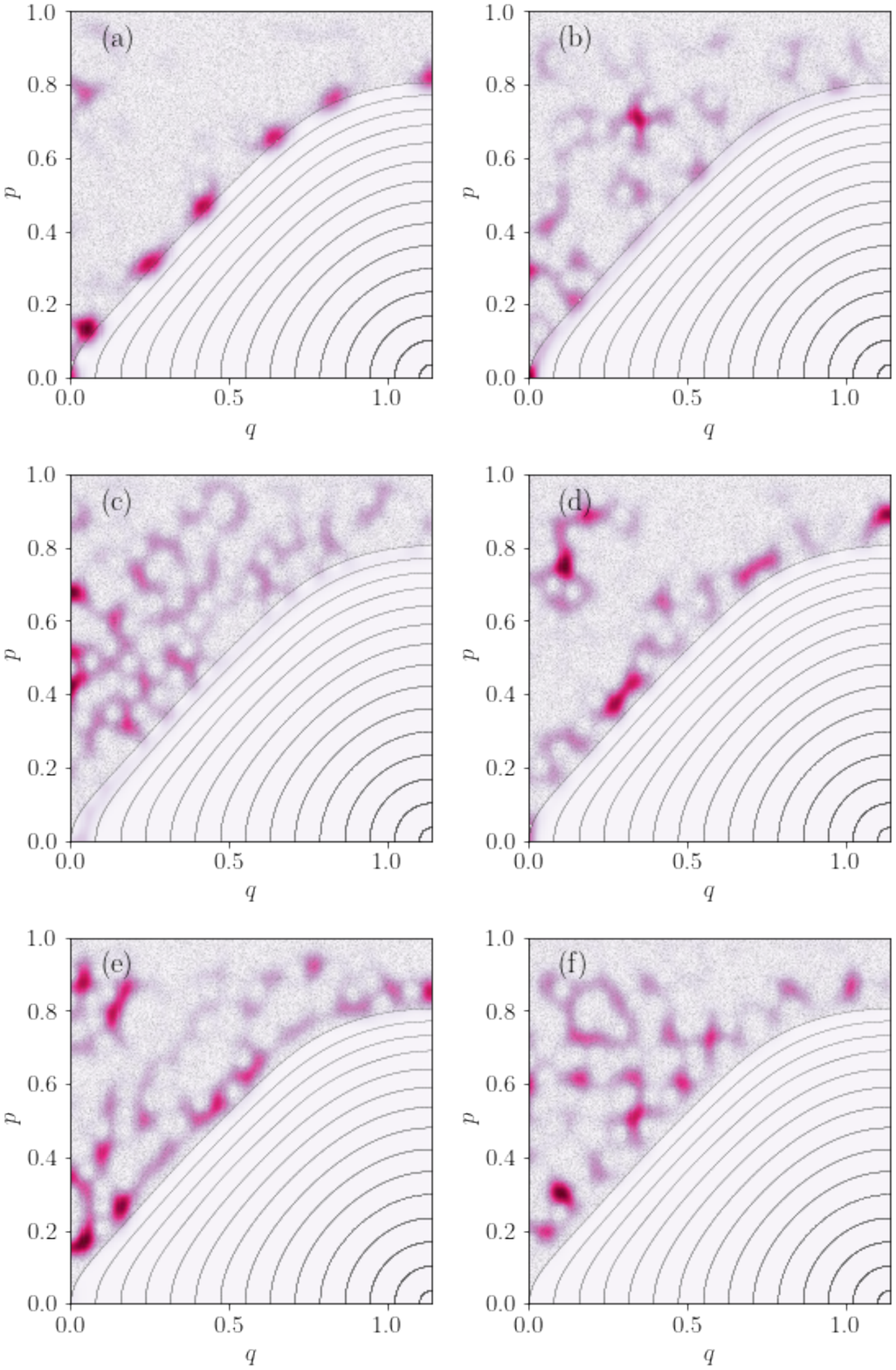}
   \par\end{centering}
 \caption{The billiard $B=0.42$, typical chaotic PH functions.
 We show only 1/8 of the phase space, due to the symmetries,
 for the even-even parity. The data $(k,A)$ are as follows from
 (a) to (f): (1220.944971, 0.220), (1321.944906, 0.323), (1245.390067, 0.371),
 (1215.916288, 0.269), (1274.016160, 0.294), (1293.913153, 0.296).}
\label{fig11}
\end{figure}

\begin{figure}
 \begin{centering}
    \includegraphics[width=9cm]{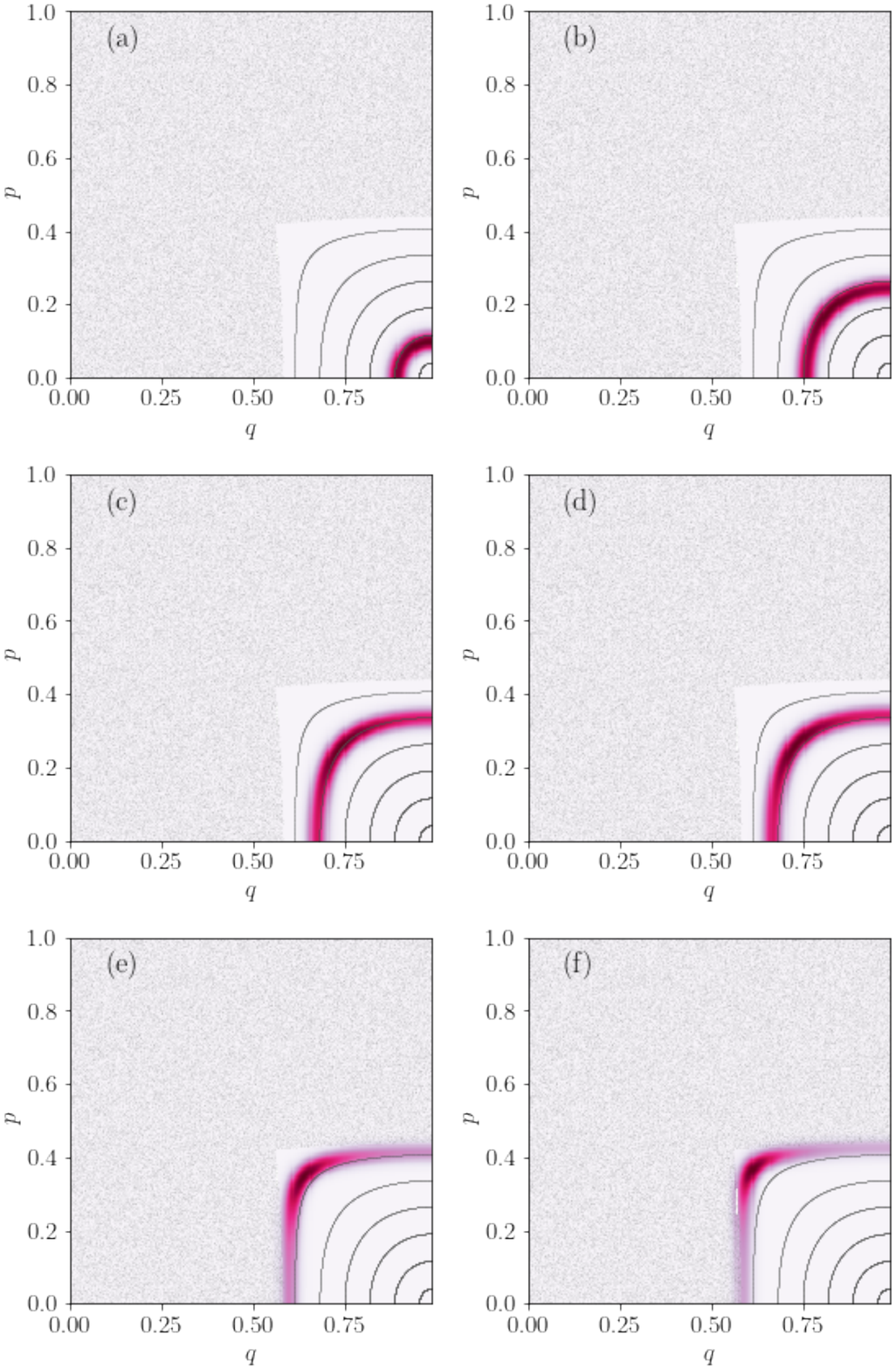}
   \par\end{centering}
 \caption{The billiard $B=0.55$, typical regular PH functions.
   We show only 1/8 of the phase space, due to the symmetries,
   for the even-even parity. The data $(k,A)$ are as follows from
   (a) to (f): (1305.660306, 0.013), (1243.133189, 0.032),
   (1291.780651, 0.04), (1228.988763, 0.046), (1332.681097,
   0.056), (1345.816393, 0.057).}
\label{fig12}
\end{figure}

\begin{figure}
 \begin{centering}
    \includegraphics[width=9cm]{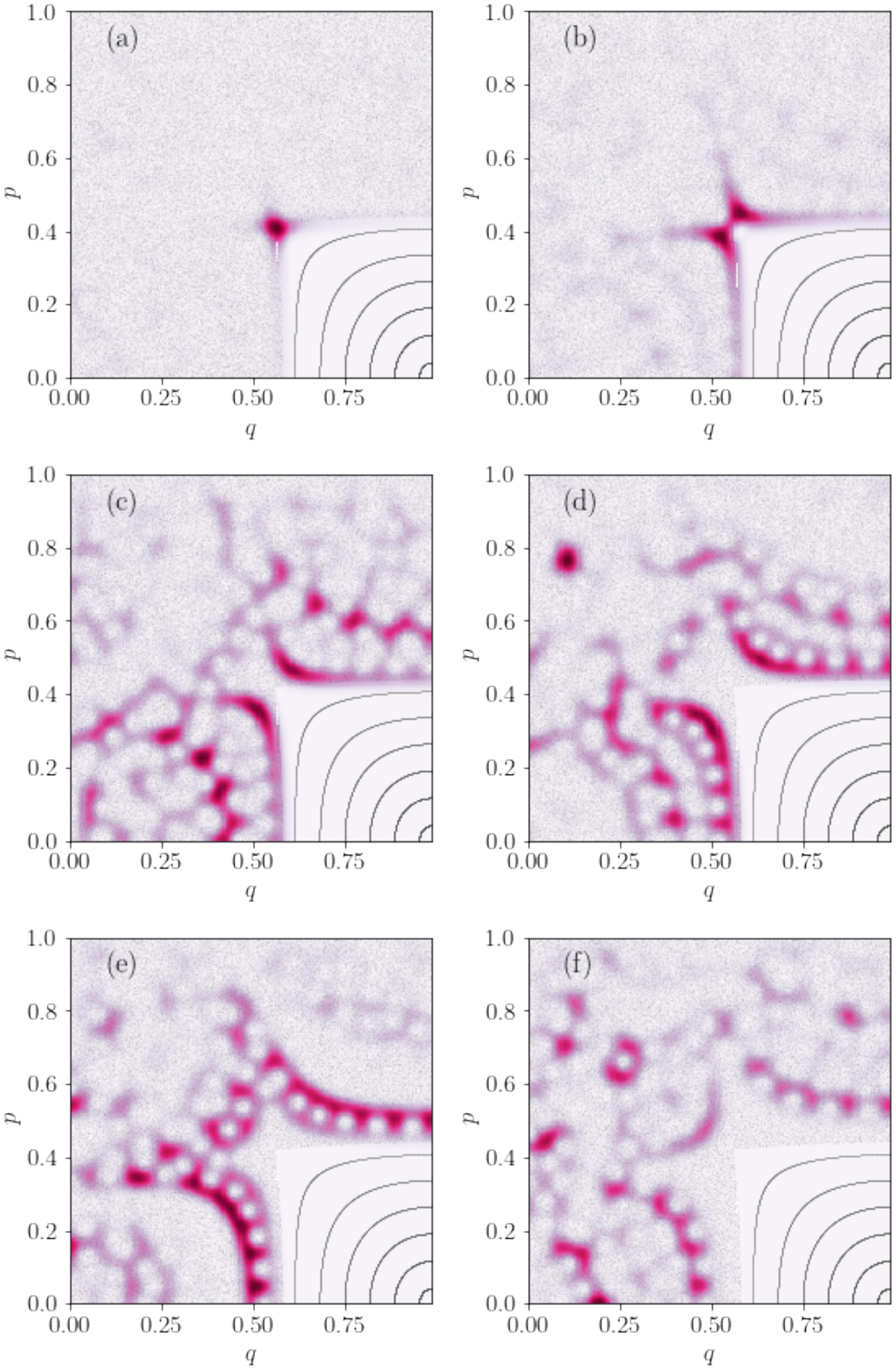}
   \par\end{centering}
 \caption{The billiard $B=0.55$, typical chaotic PH functions.
 We show only 1/8 of the phase space, due to the symmetries,
 for the even-even parity. The data $(k,A)$ are as follows from
 (a) to (f): (1345.185436, 0.304), (1351.819714, 0.382), (1358.262474, 0.520),
 (1298.492007, 0.446), (1312.074256, 0.454), (1377.733122, 0.481).}
\label{fig13}
\end{figure}

\begin{figure}
 \begin{centering}
    \includegraphics[width=9cm]{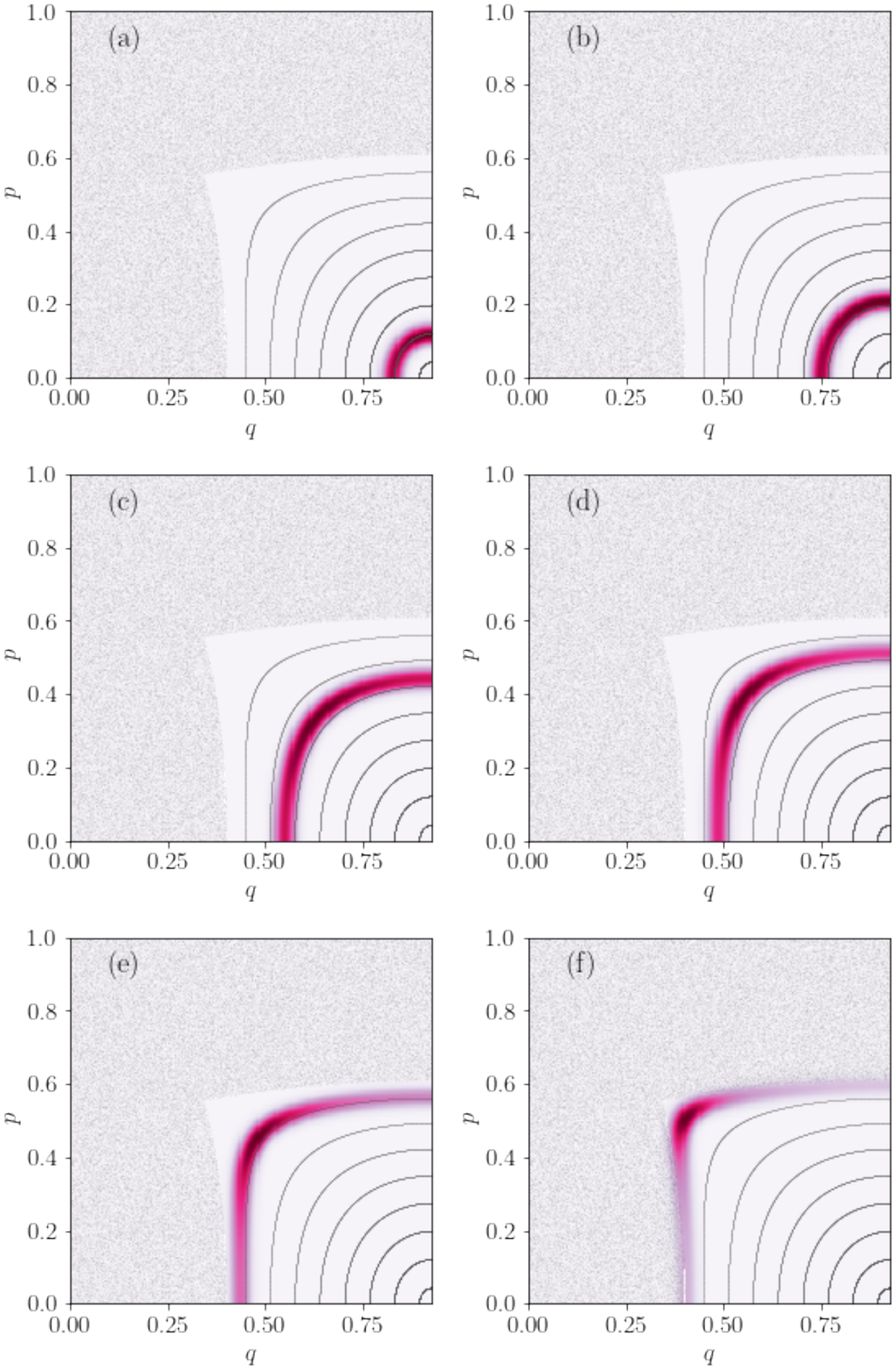}
   \par\end{centering}
 \caption{The billiard $B=0.6$, typical regular PH functions.
   We show only 1/8 of the phase space, due to the symmetries,
   for the even-even parity. The data $(k,A)$ are as follows from
   (a) to (f): (1377.224616, 0.015), (1392.401438, 0.026),
   (1242.886408, 0.060), (1225.928590, 0.072),
   (1369.637765, 0.077), (1208.778186, 0.085).}
\label{fig14}
\end{figure}

\begin{figure}
 \begin{centering}
    \includegraphics[width=9cm]{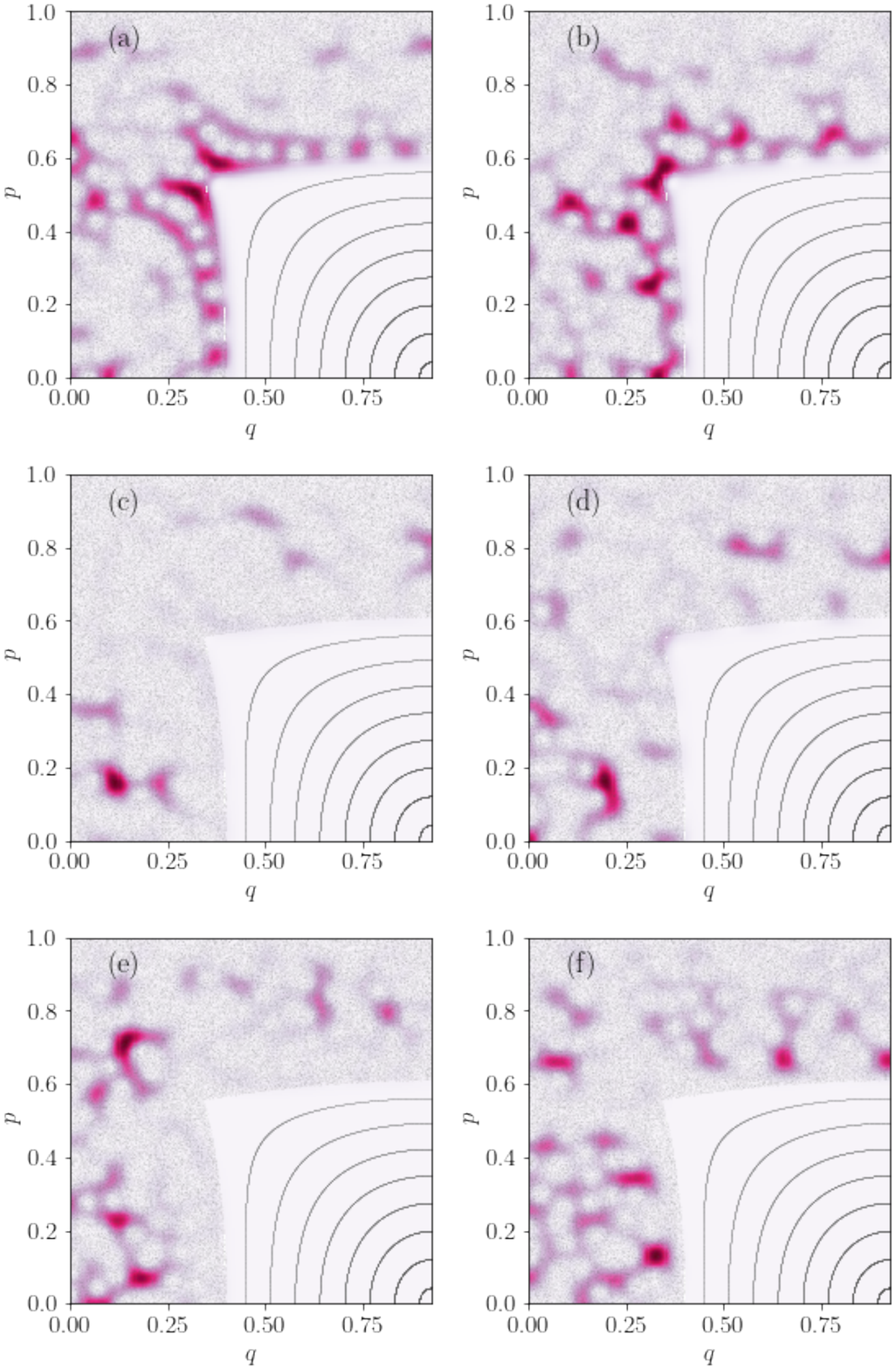}
   \par\end{centering}
 \caption{The billiard $B=0.6$, typical chaotic PH functions.
 We show only 1/8 of the phase space, due to the symmetries,
 for the even-even parity. The data $(k,A)$ are as follows from
 (a) to (f): (1315.935075, 0.406), (1323.828031, 0.406),
 (1307.989036, 0.329), (1267.431005, 0.383), (1370.279271, 0.364),
 (1275.624529, 0.392).}
\label{fig15}
\end{figure}

In Figs. \ref{fig10} - \ref{fig11}
we show examples of PH functions for the quarter billiard
$B=0.42$ of even-even symmetry for six typical regular eigenstates and
six typical chaotic eigenstates.
Due to the double reflection symmetry and time reversal symmetry,
we show only 1/8 of the phase space. The intensity of the PH functions
is encoded in red, while the underlying classical phase space
is plotted in grey, namely the invariant tori in the regular island, while
uniformly grey region represents the chaotic sea.
It is seen that regular PH functions  are very well
collapsed on the invariant tori, while the chaotic ones "live" in
the classically chaotic region and are localized to some extent.
The degree of localization will be quantified in the next Sec. \ref{sec5}
in terms of the entropy localization measure $A$. The values of
$A$ in captions of Figs. \ref{fig10} - \ref{fig15} are not rescaled,
not divided by $\chi_c$ - see next Sec. \ref{sec5}.

In Figs. \ref{fig12}-\ref{fig13} and \ref{fig14}-\ref{fig15}
we show analogous results for the billiards $B=0.55,\;0.6$,
respectively.


\section{The statistical properties of the entropy localization
  measure of PH functions}
\label{sec5}

As is well known, there are at least three localization measures of
chaotic eigenstates, namely the entropy localization measure $A$,
the correlation localization measure $C$, and the normalized
inverse participation ratio $R=nIPR$. They have been found to be
linearly related and equivalent
\cite{BatRob2013A,BatRob2013B,BLR2018,BLR2019B,BLR2020}

The {\em entropy localization measure} of a {\em single
eigenstate}  $H_m(q,p)$, denoted by $A_m$ is defined as

\be \label{locA}
A_m = \frac{\exp I_m}{N_c},
\ee
where

\be  \label{entropy}
I_m = - \int dq\, dp \,H_m(q,p) \ln \left((2\pi\hbar)^f H_m(q,p)\right)
\ee
is the information entropy.  Here $f$ is the number of degrees
of freedom (for 2D billiards $f=2$, and for surface of section it is
$f=1$) and $N_c$ is a number of cells on the 
classical chaotic domain, $N_c=\Omega_c/(2\pi\hbar)^f$, where
$\Omega_c$ is the classical phase space volume of
the classical chaotic component. In the case of the
uniform distribution (extended eigenstates) $H=1/\Omega_C={\rm const.}$
the localization measure is $A=1$, while in the case
of the strongest localization $I=0$, and $A=1/N_C \approx 0$.
The Poincar\'e-Husimi function $H(q,p)$
(\ref{Husfun}) (normalized) was calculated on the grid points $(i,j)$
in the phase space $(s,p)$,  and
we express the localization measure in terms of the discretized function.
In our numerical calculations we have put $2\pi\hbar=1$, and
thus we have $H_{ij}=1/N$ in the case of complete extendedness.
$N$ is the number of grid points
of the rectangular mesh with cells of equal area.
In case of maximal localization
we have $H_{ij}=1$ at just one point, and zero elsewhere.
In all calculations we
have used the grid of $200\times 400$ points, thus $N = 80000$.

As is well known the localization measures of a number
of consecutive eigenstates over a certain energy interval
display a distribution  $P(A)$. In classically
ergodic systems with no stickiness $A$ obeys the beta distribution
\cite{BLR2020,WR2020},
while in mixed-type systems \cite{BLR2019B} it has a
nonuniversal distribution with typically two peaks, as well as also
in ergodic systems with strong stickiness \cite{LLR2020}.

The {\em beta distribution is}

\be  \label{betadistr}
P(A) = C A^a (A_0-A)^b,
\ee
where $A_0$ is the upper limit of the interval $[0,A_0]$ on
which $P(A)$ is defined, and the two exponents $a$ and $b$
are positive  real numbers, while $C$ is the normalization constant
such that $\int_0^{A_0} P(A)\,dA = 1$, i.e.

\be \label{C}
C^{-1} = A_0^{a+b+1} B(a+1,b+1),
\ee
where $B(x,y) = \int_0^1 t^{x-1} (1-t)^{y-1} dt$ is the beta function.

Thus we have for the first moment

\be \label{mA}
\mA = A_0 \frac{a+1}{a+b+3},
\ee
and for the second moment

\be \label{2mA}
\left< A^2 \right>  = A_0^2 \frac{(a+2)(a+1)}{(a+b+4)(a+b+3)}
\ee
and therefore for the standard deviation  $\sigma=\sqrt{ \sA -\mA^2}$

\be \label{sigmaA}
\sigma^2  = 
A_0^2 \frac{(a+2)(b+2)}{(a+b+4)(a+b+3)^2},
\ee
such that asymptotically  $\sigma \approx A_0 \frac{\sqrt{b+2}}{a}$ when
$a\rightarrow \infty$. In this limit $P(A)$ becomes Dirac delta
function peaked at $A=A_0$,  $P(A)=\delta (A_0-A)$.

The maximal value of $A_0$ is typically empirically
$A_0 \approx 0.7$. The random
wavefunction model yields $A_0=0.694$  \cite{LozejThesis}.
It cannot be $1$, as PH function
never is uniformly constant, but oscillates, and in the most chaotic
random case reaches the said value $0.694$. If we
compare $P(A)$ between various systems with various sizes $\chi_c$ of
the chaotic component on the phase portrait, we must divide (normalize)
the actually calculated $A$ by $\chi_c$.

\begin{figure}
 \begin{centering}
    \includegraphics[width=9cm]{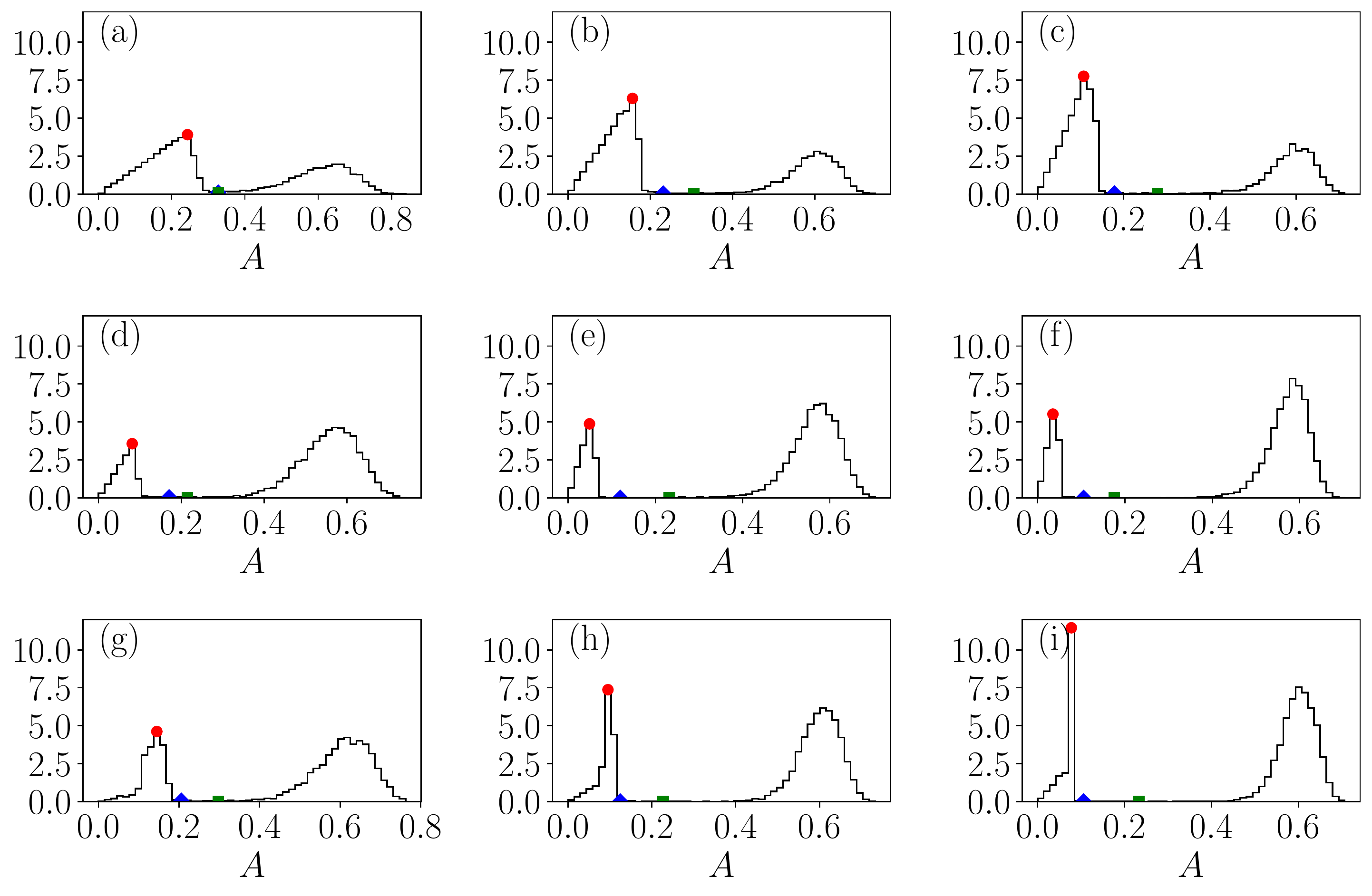}
   \par\end{centering}
 \caption{The histograms $P(A)$ for three billiards, $B=0.42$ (first row),
   $B=0.55$ (second row), and $B=0.6$ (third row), at various energies:
   (a,d,g) $k_0=640$, (b,e,h) $k_0=1760$, and (c,f,i) $k_0=2880$.
   The red dot denotes the  maximum of the first peak at $A_c$, the blue
   diamond is the point of first increase after cut-off, and
   the green square denotes the minimum on the plateau between the two peaks.
 All states are of odd-odd parity.}
\label{fig16}
\end{figure}

We have done this for each of the three billiards $B=0.42,\;0.5,\;0.6$ for
all nine different values of $k_0$.  In Fig. \ref{fig16} we show the $P(A)$
histograms for $k_0=640,\;1760,\;2880$ for the odd-odd symmetry class.
We observe that $A_0$ is indeed close to $0.7$ in all cases.
Furthermore, there are two major peaks. The left one and the right one,
with an almost zero-level plateau between them. We shall see that the
left peak at smaller $A$
comprises the regular eigenstates, while the right one the chaotic
ones. In all cases $P(A)$ rises from $A=0$ linearly up to a maximum
at $A_c$ (denoted by a thick red dot), then displays a sharp
(almost discontinuous cut-off) drop down to zero, rises again 
(denoted by a thick blue diamond), reaches the minimum
between the two peaks (denoted by a thick green square) and then rises
again forming the second peak comprising the chaotic states.
The values of the  data $(A_c,A_0)$ from (a) to (i) are:
(0.235, 0.839), (0.149, 0.747), (0.100, 0.713), (0.074, 0.742),
(0.042, 0.704), (0.028, 0.703), (0.137, 0.763), (0.088, 0.733),
(0.071, 0.707).

\begin{figure}
 \begin{centering}
    \includegraphics[width=9cm]{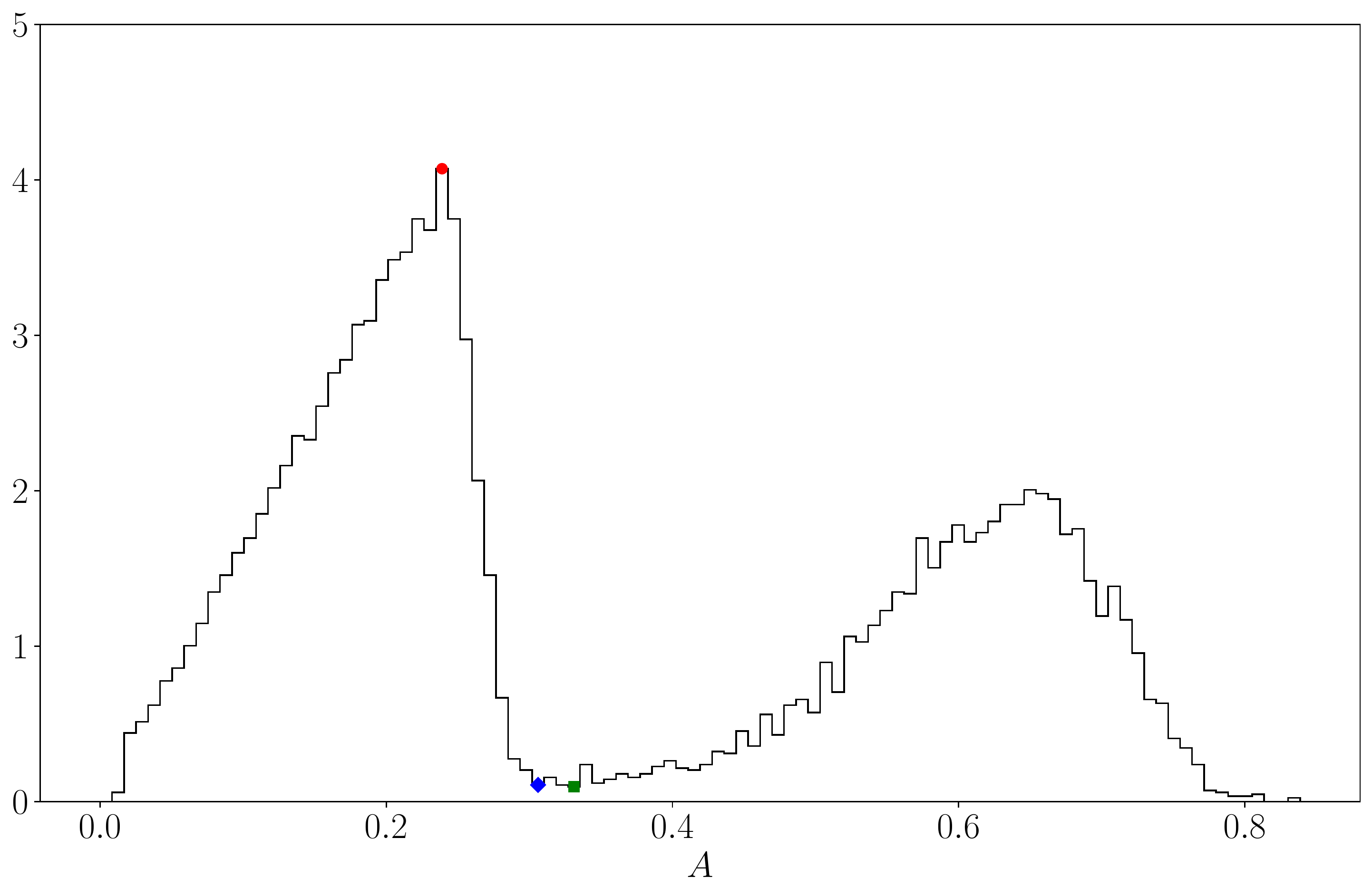}
   \par\end{centering}
 \caption{The enlarged histogram $P(A)$ for the billiard $B=0.42$ from
   Fig. \ref{fig16}(a).
   The red dot denotes the cut-off at the
   maximum of the first peak at $A_c=0.235$, the blue diamond denotes the 
   rise again, and  the green square denotes the minimum on the
   plateau between the two peaks. The maximal $A$ is $A_0=0.839$.
 All states are of odd-odd parity.}
\label{fig17}
\end{figure}

This structure is quite well understood. In order to make the details more
visible we show in Fig. \ref{fig17}
the enlarged histogram of Fig. \ref{fig16}(a). The linear
rise of the regular peak is understood as follows. Since the thickness
of the PH functions on the invariant tori (see Figs. \ref{fig10}, \ref{fig12}
and \ref{fig14})
is constant, their effective area and thus $A$
is proportional to the action $I$ or the length
of the underlying invariant tori.
The number ${\cal N}$
of eigenstates up to the given action $I$ is proportional to
the area inside the regular region encircled by the given invariant torus
of action $I$, thus ${\cal N} (I) \propto I^2$. Consequently,
${\cal N} (I) \propto A^2$, and its density $P(A) = d{\cal N}/dA \propto A$.
The cut-off at $A_c$ (estimated by the value of $A$ at the
maximum of the first peak) is well understood:
it corresponds to the boundary of the regular region (the bordering,
 last invariant torus).

\begin{figure}
 \begin{centering}
    \includegraphics[width=9cm]{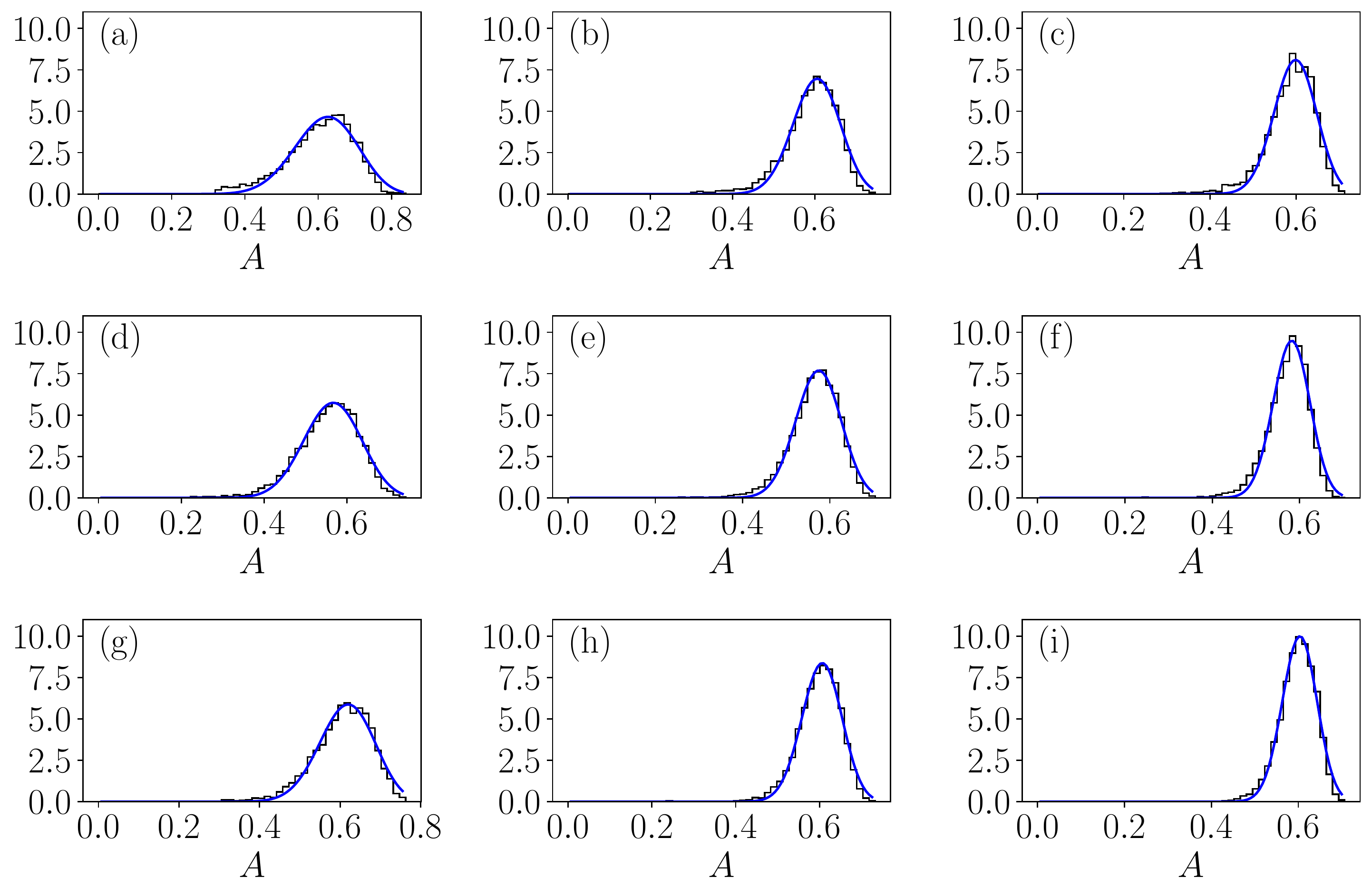}
   \par\end{centering}
 \caption{The histograms of the chaotic peak 
   $P(A)$ of Fig. \ref{fig16} for the billiards $B=0.42$ (first row),
   $B=0.55$ (second row), and $B=0.6$ (third row), at various energies:
   (a,d,g) $k_0=640$, (b,e,h) $k_0=1760$, and (c,f,i) $k_0=2880$.
   All states are of odd-odd parity. The agreement
   with the best fitting beta distribution is obvious. The parameters
   $(a,b)$ of the beta distribution (\ref{betadistr}) are, from
   (a) to (i): (20.100, 12.380), (44.209, 29.091), (59.477, 40.172),
   (28.982, 22.380), (52.361, 39.027), (80.315, 57.947),
   (31.734, 19.835), (63.872, 41.757), (90.748, 59.885). }
\label{fig18}
\end{figure}

The regular and chaotic peaks can be separated. When this is done,
we find an excellent agreement of the second peak (after normalization)
with the beta distribution, like for ergodic systems without stickiness
regions, demonstrated in Fig. \ref{fig18}.

Next we would like to understand and quantify how the histograms $P(A)$
presented in Figs. \ref{fig16} tend to their semiclassical values as
$k_0 \rightarrow \infty$. The limiting form must be as follows

\be \label{limitPA}
P(A) = \rho_1 \delta(A) + \rho_2 \delta(A_0-A).
\ee
where $\rho_1+\rho_2=1$ are the two classical parameters. For the cases
of Fig. \ref{fig16} we can separate the left and right peak
(at the green point) and then calculate the ratio of the number
of regular levels and the total number of levels in the histogram,
which must agree with the classical parameter $\rho_1$.
We find for the first row ($B=0.42$) $\rho_1=$  $0.5886$,
$0.6032$ and $0.6102$, compared
to the classical value $\rho_1=0.5873$, the second
row ($B=0.55$) $\rho_1=$ $0.1924$, $0.1808$ and $0.1778$,
compared to the classical value
$\rho_1=0.1924$, and the third row ($B=0.6$) $\rho_1=$
$0.2912$, $0.2485$ and $0.2433$,
compared to the classical value $\rho_1=0.3662$,
which is quite satisfactory, and in agreement with the Berry-Robnik
picture, except for the cases in the third row.

\begin{figure}
 \begin{centering}
    \includegraphics[width=9cm]{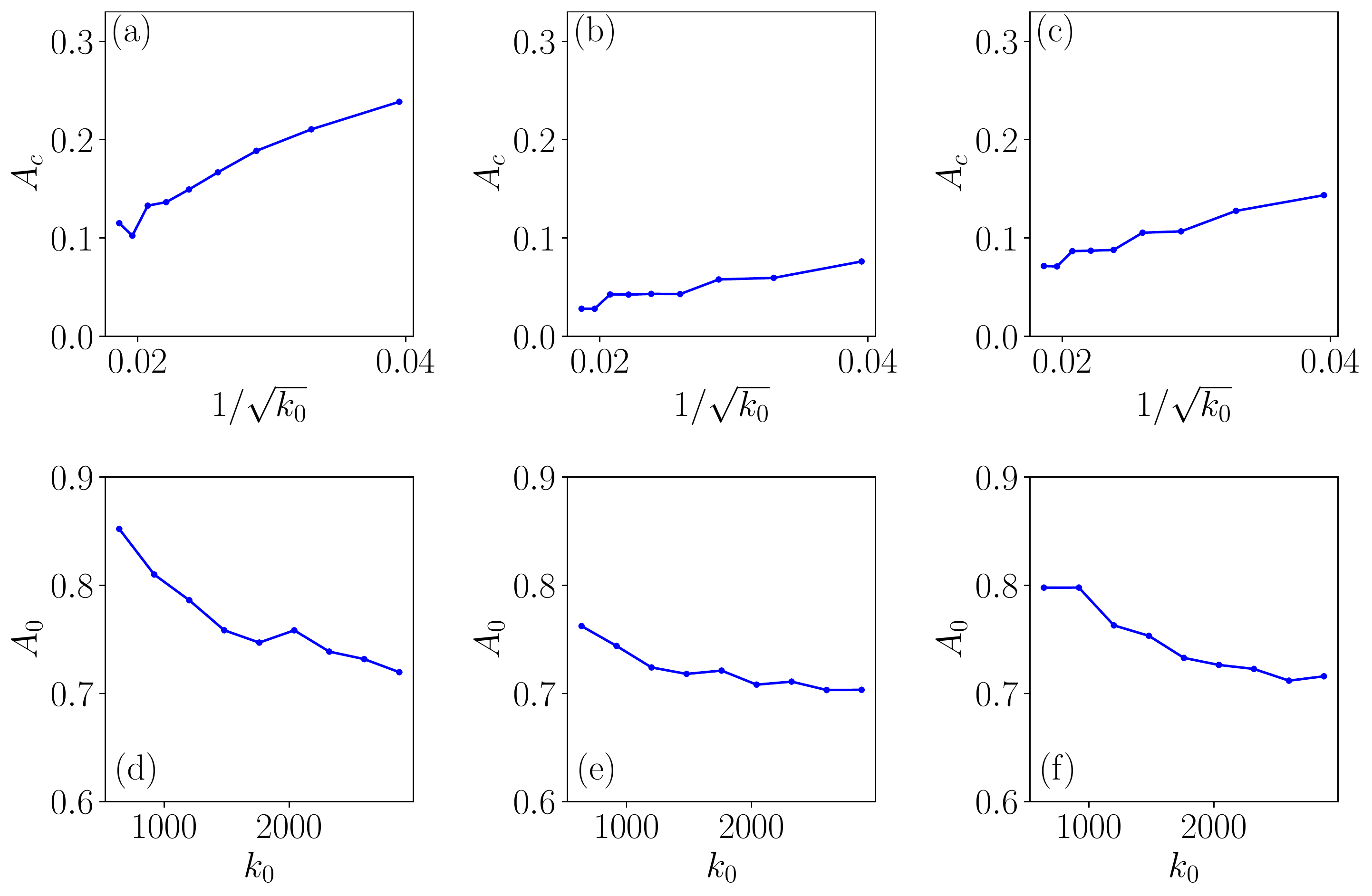}
   \par\end{centering}
 \caption{Top row: The cut-off value $A_c$ as a function of $1/\sqrt{k_0}$ for
   $B=0.42,\;0.55,\;0.6$ in (a,b,c), respectively. $A_c$ tends to zero
   with increasing $k_0$, linearly with $1/\sqrt{k_0}$.
   Bottom row: $A_0$    as a function of $k_0$ for  $B=0.42,\;0.55,\;0.6$
   in (d,e,f), respectively. $A_0$ tends to its limiting value
   $\approx 0.7$ with $k_0\rightarrow \infty$. Each data point refers to
 the distributions of $P(A)$ of all four parities.}
\label{fig19}
\end{figure}

Finally, we would like to analyze the behaviour of $A_c$ and $A_0$ as
functions of the energy, or $k_0$. Looking at the regular PH functions
in Figs. \ref{fig10}, \ref{fig12} and \ref{fig14}, we see that each
PH function is uniformly spread over the invariant torus, according
to PUSC, and this remains so in the (deep) semiclassical limit. Therefore
the entropy localization measure $A$, which also is the effective
area of the PH function, decreases linearly with the thickness
of the PH function. The latter one decreases linearly with the
square root of the Planck constant $2\pi\hbar$, like in the scars
of Heller \cite{Heller1984}. In billiards described by
the Helmholtz equation (\ref{Helmholtz}), using our units,
the effective Planck constant is $1/k_0$, and thus $A_c$
should scale as $A_c \propto 1/\sqrt{k_0}$. A similar argument
leads to the conclusion that $A_0$ as a function of $k_0$ should
tend asymptotically with $k_0\rightarrow \infty$
to the maximum value $A_0=0.694 \approx 0.7$,
corresponding to the entirely random wavefunctions.
These two observations are presented in Fig. \ref{fig19},
and are qualitatively and quantitatively clearly confirmed.

\begin{figure}
 \begin{centering}
    \includegraphics[width=9cm]{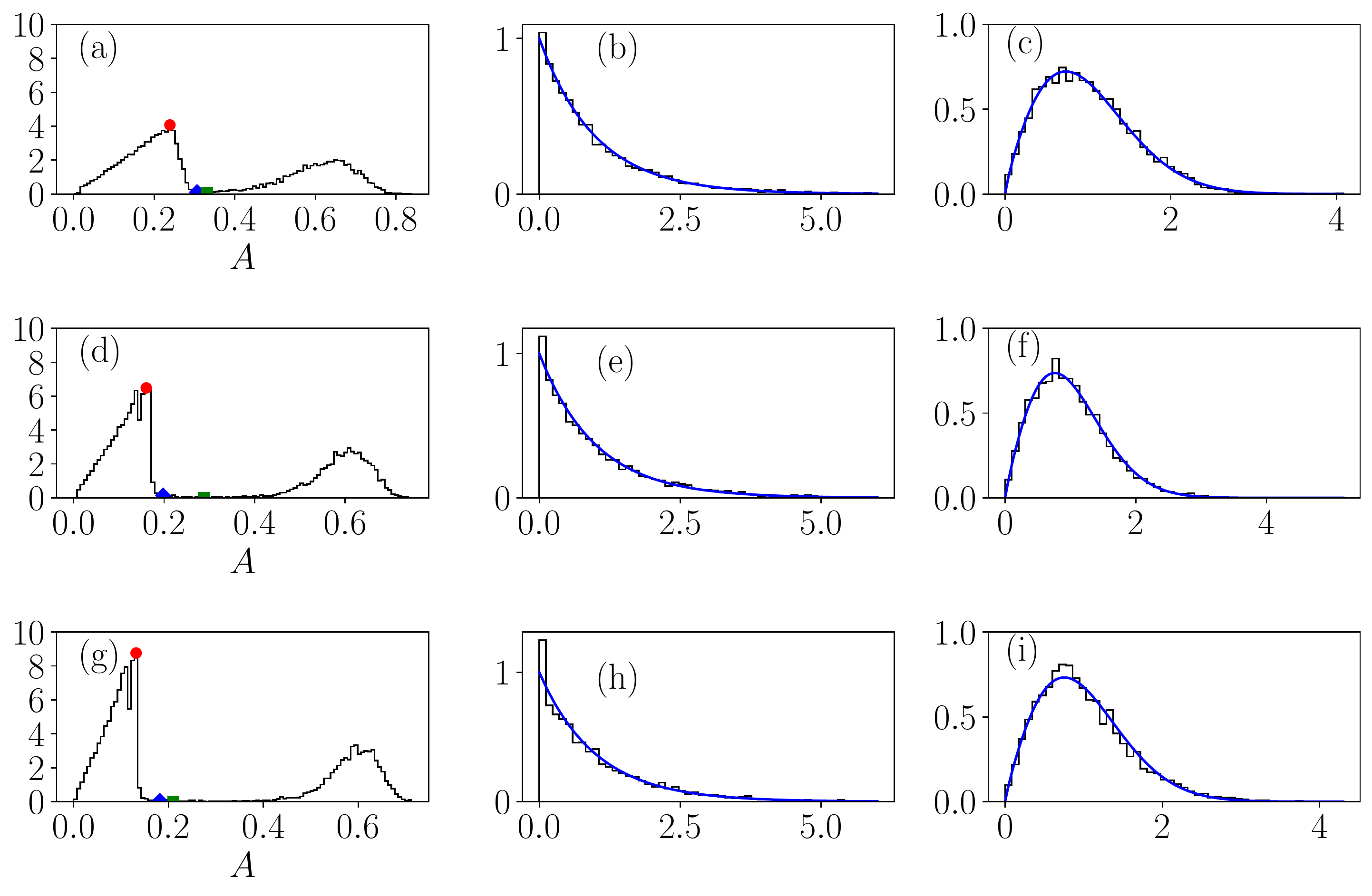}
   \par\end{centering}
 \caption{The level spacing distributions of separated regular (b,e,h)
   and chaotic levels (c,f,i),  from Fig. \ref{fig16}, $B=0.42$, at
   various energies $k_0=640$ (a,b,c), $1760$ (d,e,f) and $2880$ (g,h,i).
   The quantum $\rho_1$ in (b,e,h) is $0.5886$, $0.6032$ and $0.6102$,
   to be compared to the classical value $0.5873$. The parameter values
   $\beta$ are (c,f,i): $0.8344$, $0.8991$ and $0.8803$, which are
   somewhat smaller than in Fig. \ref{fig4}, but compatible with Fig.
   \ref{fig7}. The full curves are the Poisson distribution and the
   best fitting Brody distribution. All states are of odd-odd parity.}
\label{fig20}
\end{figure}

\begin{figure}
 \begin{centering}
    \includegraphics[width=9cm]{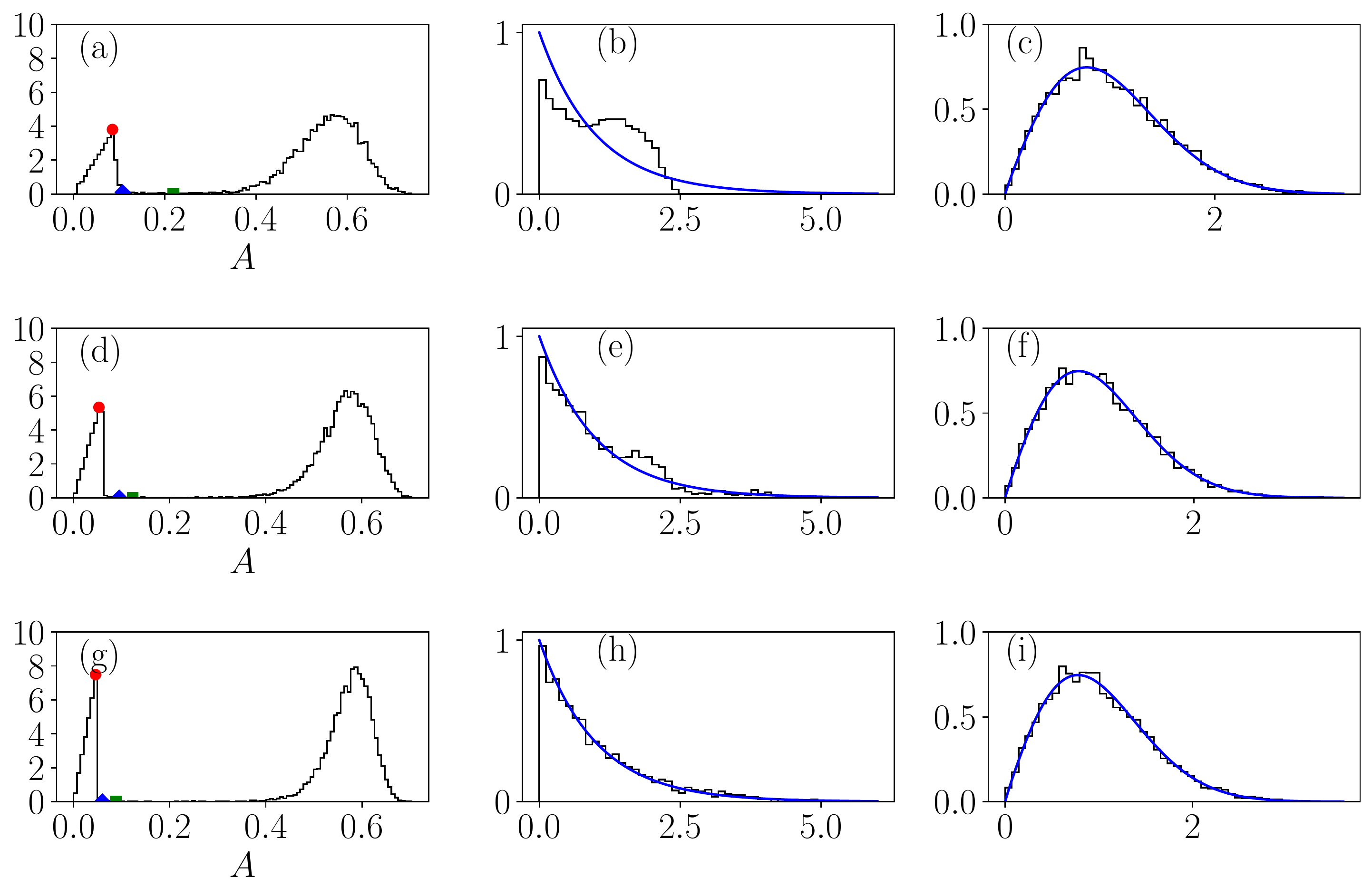}
   \par\end{centering}
 \caption{ The level spacing distributions of separated regular (b,e,h)
   and chaotic levels (c,f,i),  from Fig. \ref{fig16}, $B=0.55$, at
   various energies $k_0=640$ (a,b,c), $1760$ (d,e,f) and $2880$ (g,h,i).
   The quantum $\rho_1$ in (b,e,h) is $0.1924$, $0.1808$ and $0.1778$,
   to be compared to the classical value $0.1924$. The parameter values
   $\beta$ are (c,f,i): $0.9447$, $0.9466$ and $0.9439$, which are
   very close to the value in Fig. \ref{fig5}, and in agreement with Fig.
   \ref{fig8}.  The full curves are the Poisson distribution and the
   best fitting Brody distribution. All states are of odd-odd parity.}
\label{fig21}
\end{figure}

\begin{figure}
 \begin{centering}
    \includegraphics[width=9cm]{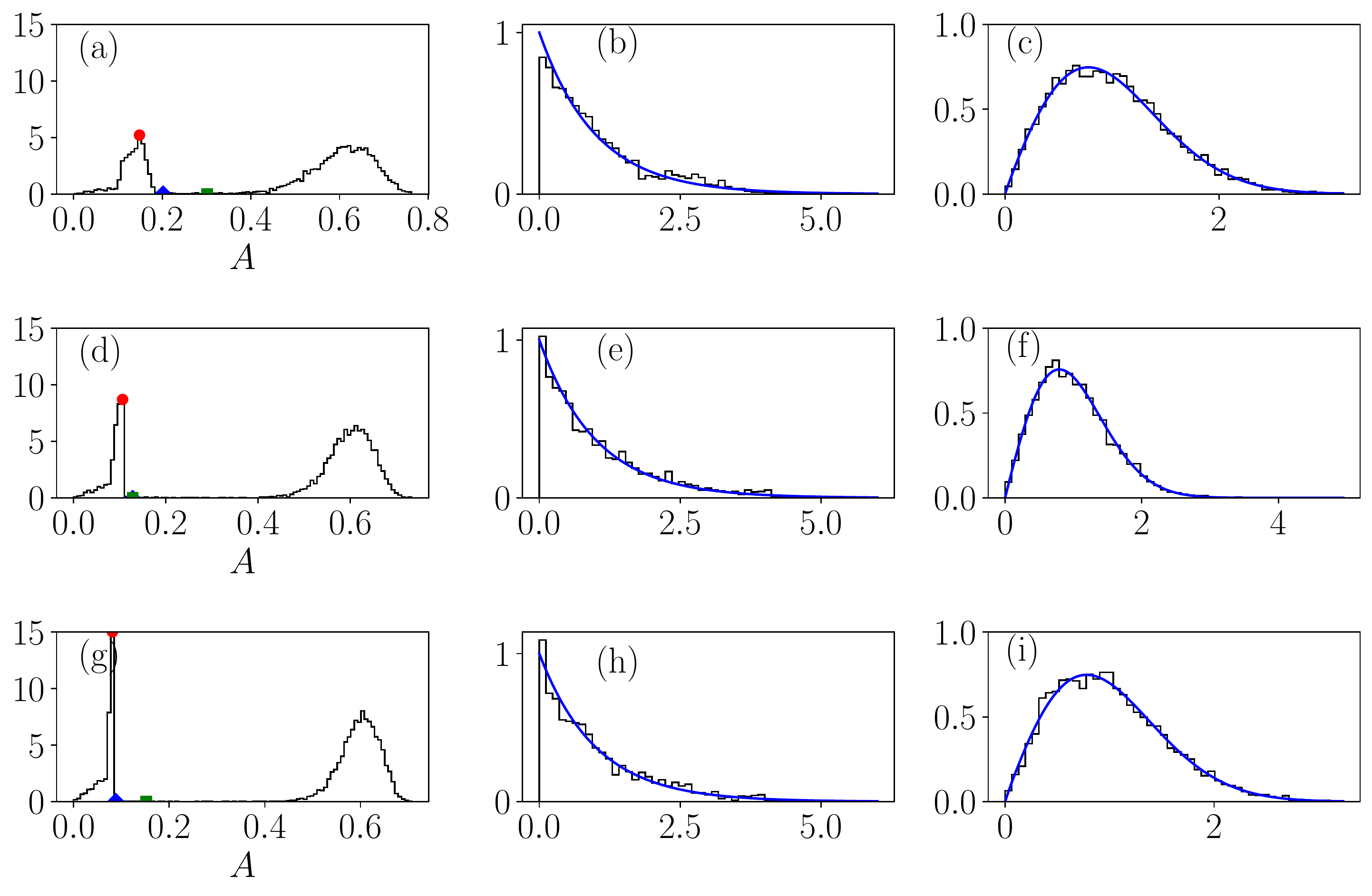}
   \par\end{centering}
 \caption{ The level spacing distributions of separated regular (b,e,h)
   and chaotic levels (c,f,i),  from Fig. \ref{fig16}, $B=0.6$, at
   various energies $k_0=640$ (a,b,c), $1760$ (d,e,f) and $2880$ (g,h,i).
   The quantum $\rho_1$ in (b,e,h) is $0.2912$, $0.2485$ and $0.2433$,
   to be compared to the significantly larger
   classical value $0.3662$. The parameter values
   $\beta$ are (c,f,i): $0.9447$, $0.9466$ and $0.9439$, which are
   close to the value in Fig. \ref{fig6}, and in agreement with Fig.
   \ref{fig9}. The full curves are the Poisson distribution and the
   best fitting Brody distribution. All states are of odd-odd parity.}
\label{fig22}
\end{figure}

Based on the $P(A)$ histograms like in Fig. \ref{fig16}, we can separate the
regular and chaotic eigenstates and corresponding energy levels, the separation
point beeing taken as the green square, denoting the minimum of the
histogram on the plateau between the two peaks. The results are shown
in Figs. \ref{fig20} - \ref{fig22}. The behaviour in Figs. \ref{fig20} -
\ref{fig22} is very well in agreement with the Poisson
(except in Fig. \ref{fig21}(b)), and Brody distributions.
The latter one is close to the Wigner surmise
(\ref{WPS}), since $\beta$ is close to $1$,
and agrees with the value
obtained from the BRB distribution, as in Fig. \ref{fig5}.
All states are of odd-odd parity. These findings
corroborate the Berry-Robnik picture of separating the regular and
chaotic eigenstates  \cite{BerRob1984} in the semiclassical regime.

One comment on the separation of eigenstates based on PH
functions is necessary. Namely, there is another approach to separate
the regular and chaotic eigenstates, by looking directly on
the overlap of the PH functions with the classical regular
and chaotic regions, respectively. This has been defined and
implemented in our previous papers \cite{BatRob2013A,BatRob2013B},
see also Refs. \cite{Rob2016,Rob2020}. There we have introduced an
overlap index $M$, which in ideal case is $1$ for chaotic states
and $-1$ for regular states. In reality, in not sufficiently
deep semiclassical limit, $M$ assumes also values between $1$ and $-1$,
and the question arises, where to cut, at certain $M=M_c$.
One rather ad hoc possibility is just $M_c=0$.
But we have two possible physical criteria, (i) the classical one,
and (ii) the quantum one. In the former case we choose $M_c$
such that the fractions of regular and and chaotic states are
the classical values $\rho_1$
and $\rho_2$, respectively. The quantum criterion for $M_c$  is
such that the fit of the chaotic level spacing distribution best
agrees with the Brody distribution. This method is applicable in
a general case. It turns out that the results
do not depend strongly on the value of $M_c$, which is
quite satisfactory. However, we
found in the billiards considered in the present work
that the results are more reliable by using the
separation criterion based on $A$, in the sense that the
quality of agreement with the Poisson and Brody distributions
for the regular and chaotic levels, respectively, is better,
and the values of $\rho_1$ and $\beta$ better agree with the
values in Figs. \ref{fig4} - \ref{fig6}. Of course, such a
separation based on $A$ is only possible if the system possesses
only one regular island embedded in a homogeneous chaotic sea
with no stickiness effects, such that the regular and the chaotic
peaks in the histogram $P(A)$ are well separated. These
conditions are clearly satisfied in the lemon billiards
$B=0.42,\; 0.55,\;0.6$, so that we can take the advantage of
the separation based on $A$ against the general method based on $M$.

\section{Discussion and conclusions}
\label{sec6}

We have demonstrated in the case of three lemon billiards
$B=0.42,\; 0.55,\; 0.6$ with divided phase space but no
stickiness regions in the chaotic sea, and no significant thin
chaotic regions inside the island of regularity, that these
systems are ideal for further study of mixed-type systems classically
and quantally. Our main result is exploring and confirming the Berry-Robnik
picture of separating statistically independent regular and
chaotic eigenstates and the corresponding energy levels \cite{BerRob1984}
in the semiclassical regime. By means of Poincar\'e-Husimi (PH) functions
we have demonstrated that indeed Principle of Uniform Semiclassical
Condensation (PUSC) \cite{Rob1998} applies, as the PH functions
clearly condense either on invariant tori, or on the chaotic component.
In the latter case they can be localized as measured by the
entropy localization measure $A$, but in the strict semiclassical limit
become uniformly extended. The distribution of $A$ has two peaks,
the regular one and the chaotic one (as for their "population").
The former has a linear rise and a cut-off determined by the last
(bordering) invariant torus, while the second one exhibits the
beta distribution. In the semiclassical limit both tend to a
Dirac delta function (\ref{limitPA}). A very similar analysis has recently
been performed in  the Dicke model \cite{WR2020} describing
the coupled atomic-bosonic systems,
which has a classical corresponding Hamilton system with a smooth
potential, which is achieved by introducing the coherent states.
We believe that our present results reconfirm the evidence
that much of the described scenario is typical and universal
for Hamilton systems with the mixed-type (divided) phase space.
If the chaotic part of the system possesses additional structure,
such as strong stickiness, we find  nonuniversal
but very interesting behaviour as studied in the recent work \cite{LLR2020}.

There are important open problems, in particular to derive
by semiclassical methods the fractional power law level repulsion
$P(S)\propto S^{\beta}$ at small $S$, and the corresponding
empirically found (approximate) Brody distribution for the localized
chaotic eigenstates. Another important open theoretical problem
is to derive the beta distribution of the entropy localization
measure $A$. More work is proposed in the empirical analysis
of billiards and smooth Hamiltonian systems. One excellent candidate is
the hydrogen atom in strong magnetic field which has
important theoretical, computational, experimental and
astrophysical aspects \cite{Rob1981,Rob1982,HRW1989,WF1989,RWHG1994}.

\section{Acknowledgement}

{\em This paper is dedicated to the memory of our friend, outstanding
  scientist and organizer of science, Professor Vyacheslav
  Ivanovich Kuvshinov (1946-2020), the founder and permanent
  Editor-in-Chief of this journal, in thankfulness of his great work.}
\\\\
The support by the Slovenian Research Agency (ARRS) under
the grant J1-9112  is gratefully acknowledged.


\begin{thebibliography}{37}%
\makeatletter
\providecommand \@ifxundefined [1]{%
 \@ifx{#1\undefined}
}%
\providecommand \@ifnum [1]{%
 \ifnum #1\expandafter \@firstoftwo
 \else \expandafter \@secondoftwo
 \fi
}%
\providecommand \@ifx [1]{%
 \ifx #1\expandafter \@firstoftwo
 \else \expandafter \@secondoftwo
 \fi
}%
\providecommand \natexlab [1]{#1}%
\providecommand \enquote  [1]{``#1''}%
\providecommand \bibnamefont  [1]{#1}%
\providecommand \bibfnamefont [1]{#1}%
\providecommand \citenamefont [1]{#1}%
\providecommand \href@noop [0]{\@secondoftwo}%
\providecommand \href [0]{\begingroup \@sanitize@url \@href}%
\providecommand \@href[1]{\@@startlink{#1}\@@href}%
\providecommand \@@href[1]{\endgroup#1\@@endlink}%
\providecommand \@sanitize@url [0]{\catcode `\\12\catcode `\$12\catcode
  `\&12\catcode `\#12\catcode `\^12\catcode `\_12\catcode `\%12\relax}%
\providecommand \@@startlink[1]{}%
\providecommand \@@endlink[0]{}%
\providecommand \url  [0]{\begingroup\@sanitize@url \@url }%
\providecommand \@url [1]{\endgroup\@href {#1}{\urlprefix }}%
\providecommand \urlprefix  [0]{URL }%
\providecommand \Eprint [0]{\href }%
\providecommand \doibase [0]{http://dx.doi.org/}%
\providecommand \selectlanguage [0]{\@gobble}%
\providecommand \bibinfo  [0]{\@secondoftwo}%
\providecommand \bibfield  [0]{\@secondoftwo}%
\providecommand \translation [1]{[#1]}%
\providecommand \BibitemOpen [0]{}%
\providecommand \bibitemStop [0]{}%
\providecommand \bibitemNoStop [0]{.\EOS\space}%
\providecommand \EOS [0]{\spacefactor3000\relax}%
\providecommand \BibitemShut  [1]{\csname bibitem#1\endcsname}%
\let\auto@bib@innerbib\@empty
\bibitem [{\citenamefont {\v{C}. Lozej}\ \emph {et~al.}(2020)\citenamefont
  {\v{C}. Lozej}, \citenamefont {Lukman},\ and\ \citenamefont
  {Robnik}}]{LLR2020}%
  \BibitemOpen
  \bibfield  {author} {\bibinfo {author} {\bibnamefont {\v{C}. Lozej}},
  \bibinfo {author} {\bibfnamefont {D.}~\bibnamefont {Lukman}}, \ and\ \bibinfo
  {author} {\bibfnamefont {M.}~\bibnamefont {Robnik}},\ }\href@noop {}
  {\bibfield  {journal} {\bibinfo  {journal} {Phys. Rev. E}\ }\textbf {\bibinfo
  {volume} {103}},\ \bibinfo {pages} {012204} (\bibinfo {year}
  {2020})}\BibitemShut {NoStop}%
\bibitem [{\citenamefont {Heller}\ and\ \citenamefont
  {Tomsovic}(1993)}]{HelTom1993}%
  \BibitemOpen
  \bibfield  {author} {\bibinfo {author} {\bibfnamefont {E.~J.}\ \bibnamefont
  {Heller}}\ and\ \bibinfo {author} {\bibfnamefont {S.}~\bibnamefont
  {Tomsovic}},\ }\href@noop {} {\bibfield  {journal} {\bibinfo  {journal}
  {Phys. Today}\ }\textbf {\bibinfo {volume} {46}},\ \bibinfo {pages} {38}
  (\bibinfo {year} {1993})}\BibitemShut {NoStop}%
\bibitem [{\citenamefont {\v{C}. Lozej}(2020{\natexlab{a}})}]{Lozej2020}%
  \BibitemOpen
  \bibfield  {author} {\bibinfo {author} {\bibnamefont {\v{C}. Lozej}},\
  }\href@noop {} {\bibfield  {journal} {\bibinfo  {journal} {Phys. Rev. E}\
  }\textbf {\bibinfo {volume} {101}},\ \bibinfo {pages} {052204} (\bibinfo
  {year} {2020}{\natexlab{a}})}\BibitemShut {NoStop}%
\bibitem [{\citenamefont {St\"ockmann}(1999)}]{Stoe}%
  \BibitemOpen
  \bibfield  {author} {\bibinfo {author} {\bibfnamefont {H.-J.}\ \bibnamefont
  {St\"ockmann}},\ }\href@noop {} {\emph {\bibinfo {title} {Quantum Chaos - An
  Introduction}}}\ (\bibinfo  {publisher} {Cambridge: Cambridge University
  Press},\ \bibinfo {year} {1999})\BibitemShut {NoStop}%
\bibitem [{\citenamefont {Haake}(2001)}]{Haake}%
  \BibitemOpen
  \bibfield  {author} {\bibinfo {author} {\bibfnamefont {F.}~\bibnamefont
  {Haake}},\ }\href@noop {} {\emph {\bibinfo {title} {Quantum Signatures of
  Chaos}}}\ (\bibinfo  {publisher} {Berlin: Springer},\ \bibinfo {year}
  {2001})\BibitemShut {NoStop}%
\bibitem [{\citenamefont {Robnik}(2016)}]{Rob2016}%
  \BibitemOpen
  \bibfield  {author} {\bibinfo {author} {\bibfnamefont {M.}~\bibnamefont
  {Robnik}},\ }\href@noop {} {\bibfield  {journal} {\bibinfo  {journal} {Eur.
  Phys. J. Special Topics}\ }\textbf {\bibinfo {volume} {225}},\ \bibinfo
  {pages} {959} (\bibinfo {year} {2016})}\BibitemShut {NoStop}%
\bibitem [{\citenamefont {Robnik}(2020)}]{Rob2020}%
  \BibitemOpen
  \bibfield  {author} {\bibinfo {author} {\bibfnamefont {M.}~\bibnamefont
  {Robnik}},\ }\href@noop {} {\bibfield  {journal} {\bibinfo  {journal}
  {Nonlinear Phenomena in Complex Systems (Minsk)}\ }\textbf {\bibinfo {volume}
  {23}},\ \bibinfo {pages} {172} (\bibinfo {year} {2020})}\BibitemShut
  {NoStop}%
\bibitem [{\citenamefont {Lopac}\ \emph {et~al.}(1999)\citenamefont {Lopac},
  \citenamefont {Mrkonji\'c},\ and\ \citenamefont {Radi\'c}}]{LMR1999}%
  \BibitemOpen
  \bibfield  {author} {\bibinfo {author} {\bibfnamefont {V.}~\bibnamefont
  {Lopac}}, \bibinfo {author} {\bibfnamefont {I.}~\bibnamefont {Mrkonji\'c}}, \
  and\ \bibinfo {author} {\bibfnamefont {D.}~\bibnamefont {Radi\'c}},\
  }\href@noop {} {\bibfield  {journal} {\bibinfo  {journal} {Phys. Rev. E}\
  }\textbf {\bibinfo {volume} {59}},\ \bibinfo {pages} {303} (\bibinfo {year}
  {1999})}\BibitemShut {NoStop}%
\bibitem [{\citenamefont {Makino}\ \emph {et~al.}(2001)\citenamefont {Makino},
  \citenamefont {Harayama},\ and\ \citenamefont {Aizawa}}]{MHA2001}%
  \BibitemOpen
  \bibfield  {author} {\bibinfo {author} {\bibfnamefont {H.}~\bibnamefont
  {Makino}}, \bibinfo {author} {\bibfnamefont {T.}~\bibnamefont {Harayama}}, \
  and\ \bibinfo {author} {\bibfnamefont {Y.}~\bibnamefont {Aizawa}},\
  }\href@noop {} {\bibfield  {journal} {\bibinfo  {journal} {Phys. Rev. E}\
  }\textbf {\bibinfo {volume} {63}},\ \bibinfo {pages} {056203} (\bibinfo
  {year} {2001})}\BibitemShut {NoStop}%
\bibitem [{\citenamefont {Lopac}\ \emph {et~al.}(2001)\citenamefont {Lopac},
  \citenamefont {Mrkonji\'c},\ and\ \citenamefont {Radi\'c}}]{LMR2001}%
  \BibitemOpen
  \bibfield  {author} {\bibinfo {author} {\bibfnamefont {V.}~\bibnamefont
  {Lopac}}, \bibinfo {author} {\bibfnamefont {I.}~\bibnamefont {Mrkonji\'c}}, \
  and\ \bibinfo {author} {\bibfnamefont {D.}~\bibnamefont {Radi\'c}},\
  }\href@noop {} {\bibfield  {journal} {\bibinfo  {journal} {Phys. Rev. E}\
  }\textbf {\bibinfo {volume} {64}},\ \bibinfo {pages} {016214} (\bibinfo
  {year} {2001})}\BibitemShut {NoStop}%
\bibitem [{\citenamefont {Chen}\ \emph {et~al.}(2013)\citenamefont {Chen},
  \citenamefont {Mohr}, \citenamefont {Zhang},\ and\ \citenamefont
  {Zhang}}]{CMZZ2013}%
  \BibitemOpen
  \bibfield  {author} {\bibinfo {author} {\bibfnamefont {J.}~\bibnamefont
  {Chen}}, \bibinfo {author} {\bibfnamefont {L.}~\bibnamefont {Mohr}}, \bibinfo
  {author} {\bibfnamefont {H.-K.}\ \bibnamefont {Zhang}}, \ and\ \bibinfo
  {author} {\bibfnamefont {P.}~\bibnamefont {Zhang}},\ }\href@noop {}
  {\bibfield  {journal} {\bibinfo  {journal} {Chaos}\ }\textbf {\bibinfo
  {volume} {23}},\ \bibinfo {pages} {043137} (\bibinfo {year}
  {2013})}\BibitemShut {NoStop}%
\bibitem [{\citenamefont {Bunimovich}\ \emph {et~al.}(2016)\citenamefont
  {Bunimovich}, \citenamefont {Zhang},\ and\ \citenamefont {Zhang}}]{BZZ2016}%
  \BibitemOpen
  \bibfield  {author} {\bibinfo {author} {\bibfnamefont {L.}~\bibnamefont
  {Bunimovich}}, \bibinfo {author} {\bibfnamefont {H.-K.}\ \bibnamefont
  {Zhang}}, \ and\ \bibinfo {author} {\bibfnamefont {P.}~\bibnamefont
  {Zhang}},\ }\href@noop {} {\bibfield  {journal} {\bibinfo  {journal}
  {Communications in Math. Phys.}\ }\textbf {\bibinfo {volume} {341}},\
  \bibinfo {pages} {781} (\bibinfo {year} {2016})}\BibitemShut {NoStop}%
\bibitem [{\citenamefont {Bunimovich}\ \emph {et~al.}(2019)\citenamefont
  {Bunimovich}, \citenamefont {Casati}, \citenamefont {Prosen},\ and\
  \citenamefont {Vidmar}}]{BCPV2019}%
  \BibitemOpen
  \bibfield  {author} {\bibinfo {author} {\bibfnamefont {L.~A.}\ \bibnamefont
  {Bunimovich}}, \bibinfo {author} {\bibfnamefont {G.}~\bibnamefont {Casati}},
  \bibinfo {author} {\bibfnamefont {T.}~\bibnamefont {Prosen}}, \ and\ \bibinfo
  {author} {\bibfnamefont {G.}~\bibnamefont {Vidmar}},\ }\href@noop {}
  {\bibfield  {journal} {\bibinfo  {journal} {Experimental Mathematics}\
  }\textbf {\bibinfo {volume} {1}},\ \bibinfo {pages} {10} (\bibinfo {year}
  {2019})}\BibitemShut {NoStop}%
\bibitem [{\citenamefont {Berry}(1981)}]{Berry1981}%
  \BibitemOpen
  \bibfield  {author} {\bibinfo {author} {\bibfnamefont {M.~V.}\ \bibnamefont
  {Berry}},\ }\href@noop {} {\bibfield  {journal} {\bibinfo  {journal} {Eur. J.
  Phys.}\ }\textbf {\bibinfo {volume} {2}},\ \bibinfo {pages} {91} (\bibinfo
  {year} {1981})}\BibitemShut {NoStop}%
\bibitem [{\citenamefont {Vergini}\ and\ \citenamefont
  {Saraceno}(1995)}]{VerSar1995}%
  \BibitemOpen
  \bibfield  {author} {\bibinfo {author} {\bibfnamefont {E.}~\bibnamefont
  {Vergini}}\ and\ \bibinfo {author} {\bibfnamefont {M.}~\bibnamefont
  {Saraceno}},\ }\href@noop {} {\bibfield  {journal} {\bibinfo  {journal}
  {Phys. Rev. E}\ }\textbf {\bibinfo {volume} {52}},\ \bibinfo {pages} {2204}
  (\bibinfo {year} {1995})}\BibitemShut {NoStop}%
\bibitem [{\citenamefont {\v{C}. Lozej}(2020{\natexlab{b}})}]{LozejThesis}%
  \BibitemOpen
  \bibfield  {author} {\bibinfo {author} {\bibnamefont {\v{C}. Lozej}},\
  }\href@noop {} {\bibfield  {journal} {\bibinfo  {journal} {Ph.D. Thesis,
  University of Maribor}\ } (\bibinfo {year} {2020}{\natexlab{b}})}\BibitemShut
  {NoStop}%
\bibitem [{\citenamefont {Berry}\ and\ \citenamefont
  {Robnik}(1984)}]{BerRob1984}%
  \BibitemOpen
  \bibfield  {author} {\bibinfo {author} {\bibfnamefont {M.~V.}\ \bibnamefont
  {Berry}}\ and\ \bibinfo {author} {\bibfnamefont {M.}~\bibnamefont {Robnik}},\
  }\href@noop {} {\bibfield  {journal} {\bibinfo  {journal} {J. Phys. A: Math.
  Gen.}\ }\textbf {\bibinfo {volume} {17}},\ \bibinfo {pages} {2413} (\bibinfo
  {year} {1984})}\BibitemShut {NoStop}%
\bibitem [{\citenamefont {Brody}(1973)}]{Bro1973}%
  \BibitemOpen
  \bibfield  {author} {\bibinfo {author} {\bibfnamefont {T.~A.}\ \bibnamefont
  {Brody}},\ }\href@noop {} {\bibfield  {journal} {\bibinfo  {journal} {Lett.
  Nuovo Cimento}\ }\textbf {\bibinfo {volume} {7}},\ \bibinfo {pages} {482}
  (\bibinfo {year} {1973})}\BibitemShut {NoStop}%
\bibitem [{\citenamefont {Brody}\ \emph {et~al.}(1981)\citenamefont {Brody},
  \citenamefont {Flores}, \citenamefont {French}, \citenamefont {Mello},
  \citenamefont {Pandey},\ and\ \citenamefont {Wong}}]{Bro1981}%
  \BibitemOpen
  \bibfield  {author} {\bibinfo {author} {\bibfnamefont {T.~A.}\ \bibnamefont
  {Brody}}, \bibinfo {author} {\bibfnamefont {J.}~\bibnamefont {Flores}},
  \bibinfo {author} {\bibfnamefont {J.~B.}\ \bibnamefont {French}}, \bibinfo
  {author} {\bibfnamefont {P.~A.}\ \bibnamefont {Mello}}, \bibinfo {author}
  {\bibfnamefont {A.}~\bibnamefont {Pandey}}, \ and\ \bibinfo {author}
  {\bibfnamefont {S.~S.~M.}\ \bibnamefont {Wong}},\ }\href@noop {} {\bibfield
  {journal} {\bibinfo  {journal} {Rev. Mod. Phys.}\ }\textbf {\bibinfo {volume}
  {53}},\ \bibinfo {pages} {385} (\bibinfo {year} {1981})}\BibitemShut
  {NoStop}%
\bibitem [{\citenamefont {Batisti\'c}\ and\ \citenamefont
  {Robnik}(2010)}]{BatRob2010}%
  \BibitemOpen
  \bibfield  {author} {\bibinfo {author} {\bibfnamefont {B.}~\bibnamefont
  {Batisti\'c}}\ and\ \bibinfo {author} {\bibfnamefont {M.}~\bibnamefont
  {Robnik}},\ }\href@noop {} {\bibfield  {journal} {\bibinfo  {journal} {J.
  Phys. A: Math. Theor.}\ }\textbf {\bibinfo {volume} {43}},\ \bibinfo {pages}
  {215101} (\bibinfo {year} {2010})}\BibitemShut {NoStop}%
\bibitem [{\citenamefont {Husimi}(1940)}]{Hus1940}%
  \BibitemOpen
  \bibfield  {author} {\bibinfo {author} {\bibfnamefont {K.}~\bibnamefont
  {Husimi}},\ }\href@noop {} {\bibfield  {journal} {\bibinfo  {journal} {Proc.
  Phys. Math. Soc. Jpn.}\ }\textbf {\bibinfo {volume} {22}},\ \bibinfo {pages}
  {264} (\bibinfo {year} {1940})}\BibitemShut {NoStop}%
\bibitem [{\citenamefont {Wigner}(1932)}]{Wig1932}%
  \BibitemOpen
  \bibfield  {author} {\bibinfo {author} {\bibfnamefont {E.}~\bibnamefont
  {Wigner}},\ }\href@noop {} {\bibfield  {journal} {\bibinfo  {journal} {Phys.
  Rev.}\ }\textbf {\bibinfo {volume} {40}},\ \bibinfo {pages} {749} (\bibinfo
  {year} {1932})}\BibitemShut {NoStop}%
\bibitem [{\citenamefont {Tualle}\ and\ \citenamefont {Voros}(1995)}]{TV1995}%
  \BibitemOpen
  \bibfield  {author} {\bibinfo {author} {\bibfnamefont {J.}~\bibnamefont
  {Tualle}}\ and\ \bibinfo {author} {\bibfnamefont {A.}~\bibnamefont {Voros}},\
  }\href@noop {} {\bibfield  {journal} {\bibinfo  {journal} {Chaos Solitons
  Fractals}\ }\textbf {\bibinfo {volume} {5}},\ \bibinfo {pages} {1085}
  (\bibinfo {year} {1995})}\BibitemShut {NoStop}%
\bibitem [{\citenamefont {B\"acker}\ \emph {et~al.}(2004)\citenamefont
  {B\"acker}, \citenamefont {F\"urstberger},\ and\ \citenamefont
  {Schubert}}]{Baecker2004}%
  \BibitemOpen
  \bibfield  {author} {\bibinfo {author} {\bibfnamefont {A.}~\bibnamefont
  {B\"acker}}, \bibinfo {author} {\bibfnamefont {S.}~\bibnamefont
  {F\"urstberger}}, \ and\ \bibinfo {author} {\bibfnamefont {R.}~\bibnamefont
  {Schubert}},\ }\href@noop {} {\bibfield  {journal} {\bibinfo  {journal}
  {Phys. Rev. E}\ }\textbf {\bibinfo {volume} {70}},\ \bibinfo {pages} {036204}
  (\bibinfo {year} {2004})}\BibitemShut {NoStop}%
\bibitem [{\citenamefont {Robnik}(1998)}]{Rob1998}%
  \BibitemOpen
  \bibfield  {author} {\bibinfo {author} {\bibfnamefont {M.}~\bibnamefont
  {Robnik}},\ }\href@noop {} {\bibfield  {journal} {\bibinfo  {journal}
  {Nonlinear Phenomena in Complex Systems (Minsk)}\ }\textbf {\bibinfo {volume}
  {1}},\ \bibinfo {pages} {1} (\bibinfo {year} {1998})}\BibitemShut {NoStop}%
\bibitem [{\citenamefont {Batisti\'c}\ and\ \citenamefont
  {Robnik}(2013{\natexlab{a}})}]{BatRob2013A}%
  \BibitemOpen
  \bibfield  {author} {\bibinfo {author} {\bibfnamefont {B.}~\bibnamefont
  {Batisti\'c}}\ and\ \bibinfo {author} {\bibfnamefont {M.}~\bibnamefont
  {Robnik}},\ }\href@noop {} {\bibfield  {journal} {\bibinfo  {journal} {Phys.
  Rev. E}\ }\textbf {\bibinfo {volume} {88}},\ \bibinfo {pages} {052913}
  (\bibinfo {year} {2013}{\natexlab{a}})}\BibitemShut {NoStop}%
\bibitem [{\citenamefont {Batisti\'c}\ and\ \citenamefont
  {Robnik}(2013{\natexlab{b}})}]{BatRob2013B}%
  \BibitemOpen
  \bibfield  {author} {\bibinfo {author} {\bibfnamefont {B.}~\bibnamefont
  {Batisti\'c}}\ and\ \bibinfo {author} {\bibfnamefont {M.}~\bibnamefont
  {Robnik}},\ }\href@noop {} {\bibfield  {journal} {\bibinfo  {journal} {J.
  Phys. A: Math. Theor.}\ }\textbf {\bibinfo {volume} {46}},\ \bibinfo {pages}
  {315102} (\bibinfo {year} {2013}{\natexlab{b}})}\BibitemShut {NoStop}%
\bibitem [{\citenamefont {Batisti\'c}\ \emph {et~al.}(2018)\citenamefont
  {Batisti\'c}, \citenamefont {\v{C}. Lozej},\ and\ \citenamefont
  {Robnik}}]{BLR2018}%
  \BibitemOpen
  \bibfield  {author} {\bibinfo {author} {\bibfnamefont {B.}~\bibnamefont
  {Batisti\'c}}, \bibinfo {author} {\bibnamefont {\v{C}. Lozej}}, \ and\
  \bibinfo {author} {\bibfnamefont {M.}~\bibnamefont {Robnik}},\ }\href@noop {}
  {\bibfield  {journal} {\bibinfo  {journal} {Nonlinear Phenomena in Complex
  Systems (Minsk)}\ }\textbf {\bibinfo {volume} {21}},\ \bibinfo {pages} {225}
  (\bibinfo {year} {2018})}\BibitemShut {NoStop}%
\bibitem [{\citenamefont {Batisti\'c}\ \emph {et~al.}(2019)\citenamefont
  {Batisti\'c}, \citenamefont {\v{C}. Lozej},\ and\ \citenamefont
  {Robnik}}]{BLR2019B}%
  \BibitemOpen
  \bibfield  {author} {\bibinfo {author} {\bibfnamefont {B.}~\bibnamefont
  {Batisti\'c}}, \bibinfo {author} {\bibnamefont {\v{C}. Lozej}}, \ and\
  \bibinfo {author} {\bibfnamefont {M.}~\bibnamefont {Robnik}},\ }\href@noop {}
  {\bibfield  {journal} {\bibinfo  {journal} {Phys. Rev. E}\ }\textbf {\bibinfo
  {volume} {100}},\ \bibinfo {pages} {062208} (\bibinfo {year}
  {2019})}\BibitemShut {NoStop}%
\bibitem [{\citenamefont {Batisti\'c}\ \emph {et~al.}(2020)\citenamefont
  {Batisti\'c}, \citenamefont {\v{C}. Lozej},\ and\ \citenamefont
  {Robnik}}]{BLR2020}%
  \BibitemOpen
  \bibfield  {author} {\bibinfo {author} {\bibfnamefont {B.}~\bibnamefont
  {Batisti\'c}}, \bibinfo {author} {\bibnamefont {\v{C}. Lozej}}, \ and\
  \bibinfo {author} {\bibfnamefont {M.}~\bibnamefont {Robnik}},\ }\href@noop {}
  {\bibfield  {journal} {\bibinfo  {journal} {Nonlinear Phenomena in Complex
  Systems (Minsk)}\ }\textbf {\bibinfo {volume} {23}},\ \bibinfo {pages} {17}
  (\bibinfo {year} {2020})}\BibitemShut {NoStop}%
\bibitem [{\citenamefont {Wang}\ and\ \citenamefont {Robnik}(2020)}]{WR2020}%
  \BibitemOpen
  \bibfield  {author} {\bibinfo {author} {\bibfnamefont {Q.}~\bibnamefont
  {Wang}}\ and\ \bibinfo {author} {\bibfnamefont {M.}~\bibnamefont {Robnik}},\
  }\href@noop {} {\bibfield  {journal} {\bibinfo  {journal} {Phys. Rev. E}\
  }\textbf {\bibinfo {volume} {102}},\ \bibinfo {pages} {032212} (\bibinfo
  {year} {2020})}\BibitemShut {NoStop}%
\bibitem [{\citenamefont {Heller}(1984)}]{Heller1984}%
  \BibitemOpen
  \bibfield  {author} {\bibinfo {author} {\bibfnamefont {E.~J.}\ \bibnamefont
  {Heller}},\ }\href@noop {} {\bibfield  {journal} {\bibinfo  {journal} {Phys.
  Rev. Lett.}\ }\textbf {\bibinfo {volume} {53}},\ \bibinfo {pages} {1515}
  (\bibinfo {year} {1984})}\BibitemShut {NoStop}%
\bibitem [{\citenamefont {Robnik}(1981)}]{Rob1981}%
  \BibitemOpen
  \bibfield  {author} {\bibinfo {author} {\bibfnamefont {M.}~\bibnamefont
  {Robnik}},\ }\href@noop {} {\bibfield  {journal} {\bibinfo  {journal} {J.
  Phys. A: Math. Gen.}\ }\textbf {\bibinfo {volume} {14}},\ \bibinfo {pages}
  {3195} (\bibinfo {year} {1981})}\BibitemShut {NoStop}%
\bibitem [{\citenamefont {Robnik}(1982)}]{Rob1982}%
  \BibitemOpen
  \bibfield  {author} {\bibinfo {author} {\bibfnamefont {M.}~\bibnamefont
  {Robnik}},\ }\href@noop {} {\bibfield  {journal} {\bibinfo  {journal} {J.
  Phys. Colloque C2}\ }\textbf {\bibinfo {volume} {43}},\ \bibinfo {pages} {29}
  (\bibinfo {year} {1982})}\BibitemShut {NoStop}%
\bibitem [{\citenamefont {Hasegawa}\ \emph {et~al.}(1989)\citenamefont
  {Hasegawa}, \citenamefont {Robnik},\ and\ \citenamefont {Wunner}}]{HRW1989}%
  \BibitemOpen
  \bibfield  {author} {\bibinfo {author} {\bibfnamefont {H.}~\bibnamefont
  {Hasegawa}}, \bibinfo {author} {\bibfnamefont {M.}~\bibnamefont {Robnik}}, \
  and\ \bibinfo {author} {\bibfnamefont {G.}~\bibnamefont {Wunner}},\
  }\href@noop {} {\bibfield  {journal} {\bibinfo  {journal} {Prog. Theor. Phys.
  Suppl. (Kyoto)}\ }\textbf {\bibinfo {volume} {98}},\ \bibinfo {pages} {198}
  (\bibinfo {year} {1989})}\BibitemShut {NoStop}%
\bibitem [{\citenamefont {Wintgen}\ and\ \citenamefont
  {Friedrich}(1989)}]{WF1989}%
  \BibitemOpen
  \bibfield  {author} {\bibinfo {author} {\bibfnamefont {D.}~\bibnamefont
  {Wintgen}}\ and\ \bibinfo {author} {\bibfnamefont {H.}~\bibnamefont
  {Friedrich}},\ }\href@noop {} {\bibfield  {journal} {\bibinfo  {journal}
  {Phys. Rep.}\ }\textbf {\bibinfo {volume} {183}},\ \bibinfo {pages} {38}
  (\bibinfo {year} {1989})}\BibitemShut {NoStop}%
\bibitem [{\citenamefont {Ruder}\ \emph {et~al.}(1994)\citenamefont {Ruder},
  \citenamefont {Wunner}, \citenamefont {Herold},\ and\ \citenamefont
  {Geyer}}]{RWHG1994}%
  \BibitemOpen
  \bibfield  {author} {\bibinfo {author} {\bibfnamefont {H.}~\bibnamefont
  {Ruder}}, \bibinfo {author} {\bibfnamefont {G.}~\bibnamefont {Wunner}},
  \bibinfo {author} {\bibfnamefont {H.}~\bibnamefont {Herold}}, \ and\ \bibinfo
  {author} {\bibfnamefont {F.}~\bibnamefont {Geyer}},\ }\href@noop {} {\emph
  {\bibinfo {title} {Atoms in Strong Magnetic Fields}}}\ (\bibinfo  {publisher}
  {Heidelberg: Springer},\ \bibinfo {year} {1994})\BibitemShut {NoStop}%
\end{thebibliography}

\providecommand{\noopsort}[1]{}\providecommand{\singleletter}[1]{#1}%

\end{document}